\documentclass[aps,reprint,amsmath, amssymb,groupedaddress,floatfix]{revtex4-1}
\usepackage{natbib}
\usepackage{graphicx}
\usepackage{bm}
\usepackage{color,soul}
\usepackage{hyperref}% add hypertext capabilities
\usepackage{float}
\usepackage{color,soul}
\usepackage{algorithm2e}
\hypersetup{colorlinks=true, urlcolor=blue, citecolor=blue}
\usepackage{threeparttable}
\usepackage{mathtools}% Loads amsmath
\begin{document}

%\preprint{AIP/123-QED}

\title{Defining the Topological Influencers and Predictive Principles to Engineer Band Structure of Halide Perovskites
} 

\author{Ravi Kashikar}
\author{Mayank Gupta}
\author{B. R. K. Nanda}
\email{nandab@iitm.ac.in}
\affiliation{Condensed Matter Theory and Computational Lab, Department of Physics, \\ Indian Institute of Technology Madras, Chennai - 36, India}
\date{\today}

\date{\today}

\begin{abstract}
 Complex quantum coupling phenomena of halide perovskites are examined through ab-initio calculations and exact diagonalization of model Hamiltonians to formulate a set of fundamental guiding rules to engineer the bandgap through strain. The bandgap tuning in halides is crucial for photovoltaic applications and for establishing non-trivial electronic states. Using CsSnI$_3$ as the prototype material, we show that in the cubic phase, the bandgap reduces irrespective of the nature of strain. However, for the tetragonal phase, it reduces with tensile strain and increases with compressive strain, while the reverse is the case for the orthorhombic phase. The reduction can give rise to negative bandgap in the cubic and tetragonal phases leading to normal to topological insulator phase transition. Also, these halides tend to form a stability plateau in a space spanned by strain and octahedral rotation. In this plateau, with negligible cost to the total energy, the bandgap can be varied in a range of 1eV. Furthermore, we present a descriptor model for the perovskite to simulate their bandgap with strain and rotation. Analysis of band topology through model Hamiltonians led to the conceptualization of topological influencers that provide a quantitative measure of the contribution of each chemical bonding towards establishing a normal or topological insulator phase. On the technical aspect, we show that a four orbital based basis set (Sn-\{s,p\} for CsSnI$_3$) is sufficient to construct the model Hamiltonian which can explain the electronic structure of each polymorph of halide perovskites.
\end{abstract}

\pacs{Valid PACS appear here}% PACS, the Physics and Astronomy
                             % Classification Scheme.
\keywords{Suggested keywords}%Use showkeys class option if keyword
                              %display desired
\maketitle

\section{Introduction}

Halide perovskites, enriched with structurally modulated diverse electronic phases, are promising materials for research in the fundamental area of band topology and in the applied area of optoelectronics \cite{pv0,pv01,pv1,pv2,Nature_mat}.  These compounds with the formula ABX$_3$ (A is either organic or inorganic, B = Sn, Pb or Ge and X = Cl, Br or I) are structurally flexible to external stimuli such as pressure, temperature and chemical doping \cite{doping1,doping2,strain1,strain3,strain4,temp}. The symmetry altering structural modifications invite changes in the interplay of lattice, orbital and spin degrees of freedom to manifest trivial and non-trivial electronic phases.  Studies suggest that  if A is inorganic elements like Cs and Rb, then the centrosymmetric structure stabilizes \cite{12} and in such a situation while covalent interaction among the B and X states tune the bandgap (E$_g$), the strength of atomic spin-orbit coupling  become deterministic towards tailoring non-trivial topological phases such as topological insulator \cite{strain3,Ravi}. However, though debatable, some of the experimental and theoretical studies show that if A is an organic element like MA (CH$_3$NH$_3$) and FA (CH(NH$_2$)$_2$)the inversion symmetry breaks down and therefore both atomic SOC and Rashba SOC - that inflicts the spin-split in the momentum space - become deterministic in tailoring the electronic behavior of these quantum materials \cite{rashba_soc,rashba_soc_1}.

Specific to the studies of band topology in inorganic halide perovskites, the literature, focusing on the high temperature cubic phase, hypothesizes a pressure induced phase transition from the normal insulator (NI) to topological insulator (TI) phase via an accidental Dirac semimetal (DSM) phase \cite{Ravi,strain3}. However, experimentally halide perovskites are known to demonstrate structural polymorphs with temperature governing their stability.  With lowering in temperature  the cubic ($\alpha$) phase yields to the tetragonal($\beta$) phase and on further reduction of the temperature these compounds stabilize in the orthorhombic ($\gamma$) phase \cite{DFT_1_2, struct1,struct2}.  

Moving from $\alpha$ to $\beta$ and $\gamma$-phases via symmetry lowering introduces additional degrees of freedom in the form of bond length and bond angle, as shown in Fig. \ref{fig1}. For example, unlike the case of a cubic structure where the epitaxial strain is measured only through $c/a$, for the orthorhombic phase it is measured through $c/a$ and $b/a$ as well as through rotational and tilting angles $\theta_{ab}$ and $\theta_c$. These additional degrees of freedom in low symmetry structure introduce significant complexity to the valence and conduction bands and hence are largely avoided in the electronic structure analysis. From a positive point of view, these additional degrees offer further flexibility to tune the bandgap which is a crucial ingredient to improve the photovoltaic efficiencies as well as to open up new insight to the physics of non-trivial topology. The objective of this paper is to take into account these symmetry lowering structural distortions and possible manipulation of them through external routes such as pressure and strain  so that a set of fundamental design principles can be hypothesized for tuning the electronic structure of the halide perovskite family.

Several attempts have been made in the past to examine the band structure of halide perovskites as a function of pressure and strain. For example, Grote et al.\cite{strain1}, have studied the bandgap variation of CsPbI$_3$ and CsSnI$_3$ crystals as function of biaxial strain. However, as the SOC has not been included in this study, the non-trivial states emerging out of SOC driven band inversion cannot be established. The DFT studies carried out by Liu et al.\cite{Nano_letter} and Jin et al.\cite{strain3}, examine the uniform pressure driven NI-TI phase transition of $\alpha$- CsPbI$_3$ and CsSnI$_3$. However, the polymorphs with lower symmetry structures ($\beta$ and $\gamma$) are ignored in these studies and also the role of strain has not been explored. Several model Hamiltonian studies, primarily based on tight-binding methods, are also developed. But they are found to be inadequate to capture the mechanisms governing the band topology of the halide family in general. In some of them, the interactions are restricted to nearest neighbour A-B interaction and the crucial second neighbour B-B interactions of ABX$_3$ are ignored\cite{JPCL}. In some other even if the second neighbour interactions are considered, the anisotropic behaviour in the interaction strength emerging out of octahedral rotation and biaxial strain are not investigated\cite{strain3,Ravi}.

The universality of the electronic structure of perovskite members as explained in our earlier works \cite{Ravi, BK} enables us to examine a single compound rather than the whole family and extract the insight into the band structure with regard to the anisotropic structural distortions. In this paper, it is achieved by carrying out DFT calculations on CsSnI$_3$ polymorphs. The DFT results are served as a critical input to a minimal basis set based parametric Slater-Koster tight binding (SK-TB) Hamiltonian suited for each of the structural phases. Interestingly, we show that a minimal basis set consisting of 4 orbitals of the single pseudo-cubic lattice formed out of the B element of ABX$_3$ is sufficient to explain its band structure even for the lower symmetry structures ($\beta$ and $\gamma$). By solving the SK-TB Hamiltonian,  we propose an empirical descriptor model capable of predicting the band topology of the Halide family. The model consists of a set of linear equations [T] = [D][EC], where [T] represents the set of hopping interactions,  [D] is the family of descriptors specific to a given compound, and [EC] are the externally controlled variables such as the bond lengths and the rotational and tilting angles. 

In the context of CsSnI$_3$, we show that the bandgap can be manipulated to the tune of 1eV through epitaxial strain. Furthermore, a NI to TI phase transition can be achieved either by compressive or tensile strain for the $\alpha$-phase. However, in the $\beta$-phase such a transition is plausible only through tensile strain. In the $\gamma$-phase, the system cannot exhibit the TI state as either the bulk bandgap increases from its equilibrium value or the structure enters an unstable regime. 
\section{Structural and Computational Details}
\begin{figure}[h]
\centering
\includegraphics[angle=-0.0,origin=c,height=10.3cm,width=8.7cm]{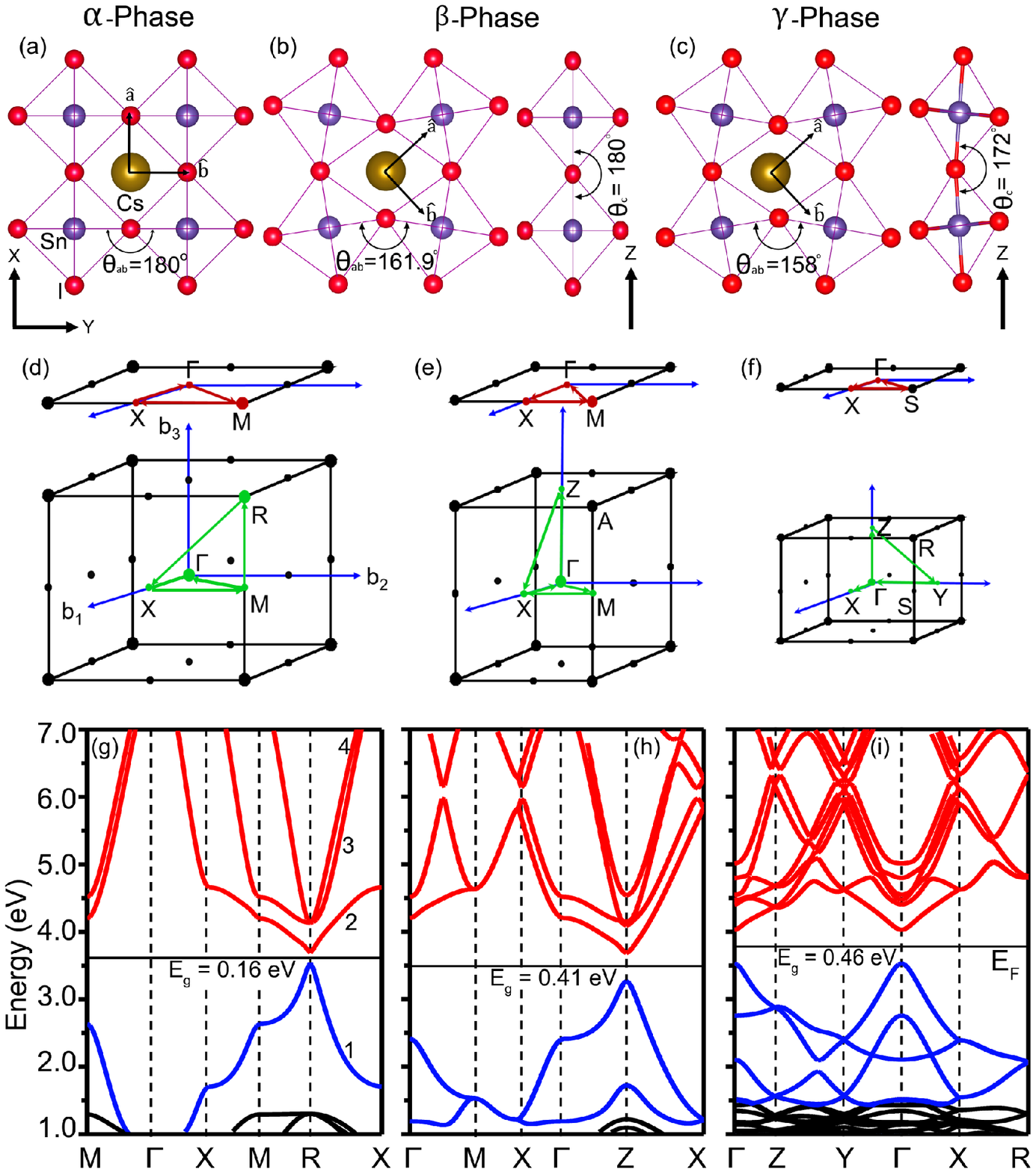}
\caption{
(a)-(c)} Crystal structure of the three polymorphs ($\alpha$: cubic, $\beta$: tetragonal, and $\gamma$: orthorhombic) of CsSnI$_3$. The $\beta$-phase arises due octahedral rotation (represented by angle $\theta_{ab}$) in the $xy$-plane.  In the $\gamma$-phase, the octahedra tilt both in the $xy$ plane and along the $z$ axis. \textcolor{red}{(d)-(f)} ) The bulk and surface Brillouin zones along with the high symmetry $k$-points of the corresponding crystal structures. The time reversal invariant momenta (TRIM) $R$ of the $\alpha$-phase is mapped to the Z and $\Gamma$ of the $\beta$ and $\gamma$-phases respectively. \textcolor{red}{(g)-(i)}  The ground state band structure corresponding to these three polymorphs, obtained using GGA+mBJ exchange correlation. The Sn-s and Sn-p dominated bands are shown in blue and red respectively. 
\label{fig1}
\end{figure}
The crystal structure of the halide perovskites in all three phases ($\alpha$, $\beta$, and $\gamma$) are shown in the  Fig. \ref{fig1}(a-c). The experimental structural parameters corresponding to CsSnI$_3$ are listed in Table I \cite{struct1,struct2,struct3}. The $\beta$-phase emerges out of the distortion of $\alpha$-phase through minor compression of $\sim$0.015\AA{} in the $ab$-plane and expansion of $\sim$0.05\AA{} along the $c$-axis. The lattice distortion is also accompanied by deviation of the angle between the neighboring octahedra in the $ab$-plane ($\theta_{ab}$) from the ideal 180$^{\circ}$. The primitive unitcell of the $\beta$-phase is formed by a $\sqrt{2}\times\sqrt{2}\times{1}$ supercell. In the $\gamma$ phase the lattice experience natural compression in all three directions. Along the $a$-axis the compression is of $\sim$0.075\AA{}, along the $b$-axis it is 0.1\AA{} and along the $c$-axis it is 0.03\AA{}. In this phase, both the angles, $\theta_{ab}$ and $\theta_c$ deviates from 180$^{\circ}$. The primitive unitcell of the $\gamma$-phase is formed out of the $\sqrt{2}\times\sqrt{2}\times2$ supercell.

The bulk and surface Brillouin zone (BZ) and the high symmetry (HS) points in the reciprocal space for each of these phases are shown in Fig. \ref{fig1}(d-f). As the primitive cell in the real space gets augmented due to lowering in the symmetry, the BZ and hence the HS points of the $\beta$ and $\gamma$-phases differ from that of the $\alpha$-phase. Consequently, The TRIM $R$ of $\alpha$-phase is mapped to Z and $\Gamma$ of $\beta$ and $\gamma$-phases respectively. 

\begin{table}[h]
\centering
\caption{Crystallographic information of CsSnI$_3$. Here, the $\theta_{ab}$ and $\theta_c$ denote the angle between  Sn-I-Sn in $xy$ plane and along $z$ direction, respectively (see Fig. \ref{fig1}). }
\begin{tabular}{ccc}
\hline
\hline
Structure,  &  Lattice & Octahedral   \\
 Glazer notation, &  parameter(\AA)& rotation\\
 Temperature range (K)&&\\
\hline
Cubic ($a_0a_0a_0$)($\alpha$)& a = 6.219 & -  \\
Pm-3m (221)&&\\
$>$425&&\\\\
Tetragonal ($a^0a^0c^+$)($\beta$)& a = 8.772,  & $\theta_{ab} =161.9^{\circ}$  \\
P4/mbm (127)        & c = 6.261   &$\theta_{c} =180^{\circ}$      \\
425-351 &&\\\\
Orthorhombic ($a^-a^-c^+$)($\gamma$)& a = 8.688, & $\theta_{ab} =158^{\circ}$   \\
Pnam (62)            & b= 8.643,  & $\theta_{c} =172.3^{\circ}$   \\  
     $<$ 351       &c = 12.378  &                               \\

\hline
\hline
\end{tabular}
\label{T1}\\
\end{table}

The electronic structure of each of the polymorphs of CsSnI$_3$, in their equilibrium and strained structures, are carried out using both density functional theory (DFT) and minimal basis set based tight-binding model Hamiltonian. For DFT, we have used the full-potential linearized augmented plane wave (FP-LAPW) method as implemented in WIEN2k simulation tool \cite{LAPW,Blaha}. The modified Becke-Johnson (mBJ) exchange potential + PBE-GGA correlation is used for the band structure calculation \cite{GGA,mbj-1,mbj-2}. However, the structural relaxation is carried out within the framework of PBE-GGA exchange-correlation (XC) functional. The basis set, employed for the SCF calculations, consists of augmented plane waves occupying the interstitial region and localized orbitals (Cs-6s, Sn-\{5s, 5p\} and I-5p ) occupying the respective muffin-tin spheres (R$_{MT}^{Cs}$ = 2.5 a.u, R$_{MT}^{Sn}$ = 2.5 a.u and R$_{MT}^{I}$ = 2.5  a.u). The number of plane waves is determined by setting $R_{MT}K_{MAX}$ to 7.0. We find that with this cutoff, the number of plane waves lies in the range 10$^3$-10$^4$. The Brillouin zone integration is carried out with a Monkhorst-Pack grid. We used a k-mesh of 10$\times$10$\times$10 (yielding 35 irreducible
points), 8$\times$8$\times$12 (yielding 60 irreducible points), and 8$\times$8$\times$6 (yielding 100 irreducible points) for the $\alpha$, $\beta$, $\gamma$-phases respectively. 
The bulk-boundary correspondence of the band topology is examined by calculating the SK-TB based surface band structure of a slab grown along (001).

\section{DFT Calculations: Results and Analysis}
\subsection{Electronic Structure of $\alpha$, $\beta$, and $\gamma$ Phases of CsSnI$_3$ }
As earlier discussed, there exist a few articles in the literature discussing the electronic structure of  CsSnI$_3$  obtained from DFT calculations \cite{PRB_struct,Lamb_1,Lamb_2,band_1,strain3,struct3}. The focus of these articles is mainly confined to examine the bandgap of the compound and its suitability for photovoltaic applications. However, the non-trivial topological aspects, where spin-orbit coupling (SOC) plays a major role, as well as bandgap modulation as a function of strain and lattice distortion which  have been largely ignored and therefore the same will be discussed in this section. 
The  Fig. \ref{fig1}(g-i) illustrates the ground state electronic structure of CsSnI$_3$ in the absence of strain which shows a monotonous increase in the bandgap as we move from the higher symmetric $\alpha$-phase to the lower symmetric $\gamma$-phase. A similar observation is also made when calculations are carried out without including the SOC \cite{PRB_struct}. Hence, it implies that this increase in bandgap with symmetry reduction is largely due to the variation in the chemical bonding.

Considering the $\alpha$-phase, for any halide and oxide perovskites ABX$_3$ we have shown that bonding and anti-bonding bands emerge  out of strong B-\{s, p\}-X-p nearest neighbor covalent  interactions.\cite{BK,Ravi}. The anti-bonding and bonding bands are separated by a set of X-p dominated non-bonding bands. Furthermore, both bonding and anti-bonding spectrum have four bands with the lower one is formed by the B-s-X-p hybridized states, and the upper three are formed by the B-p-X-p hybridized states. Also, there is a gap between the lower one and the upper three bands. In the case of halide perovskites, where the valence electron count is 20, this gap exactly lies at the Fermi level, as can be seen from Fig. \ref{fig1}, in order to make these compounds semiconducting or insulating. As the figure suggests, the $\beta$ and $\gamma$-phases, which stabilize with the distortion to the $\alpha$-phase, inherit these same salient features, but with a larger bandgap due to distortion induced reduction in the bandwidth.

Though the universality of the band structure arises from the nearest-neighbor B-\{s, p\}-X-p covalent interactions as discussed in the previous paragraph, the second neighbor B-\{s, p\}-B-\{s, p\} interactions are found to be the driving force for engineering the bandgap as well as construing non-trivial phases \cite{Ravi}. In the ground state, the second neighbor interaction strength is less than 1eV (almost one-third of nearest neighbor interaction) \cite{Ravi, BK}  which can be varied and made anisotropic easily through strain and will be discussed in the following sub-section.

\subsection{Effect of Epitaxial Strain on the Band Topology}

\begin{figure}
\centering
\includegraphics[angle=-0.0,origin=c,height=12.5cm,width=8.5cm]{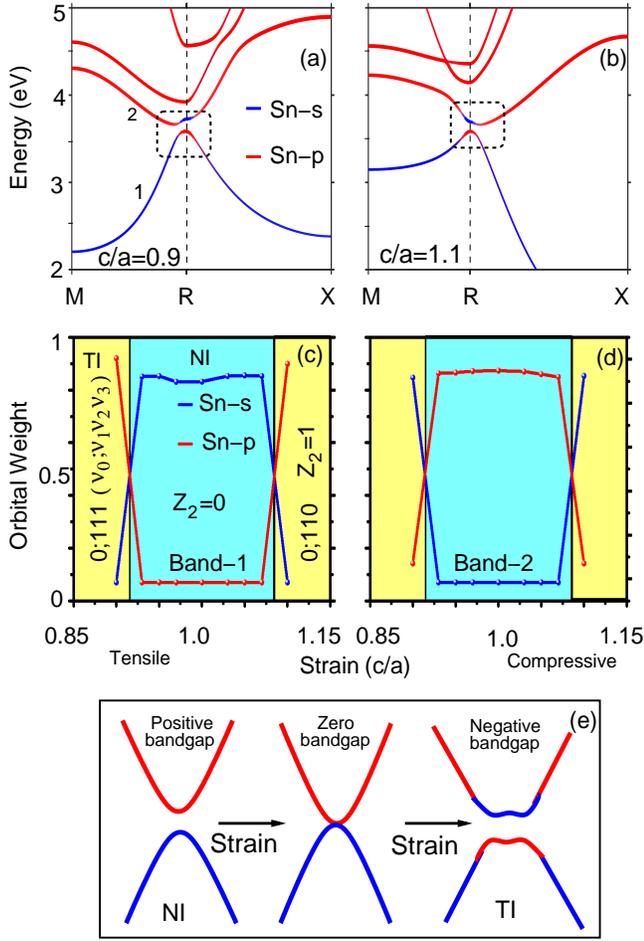}
\caption{The representative band structure of  $\alpha$-CsSnI$_3$ demonstrating s-p character inversion for band-1 and 2 under tensile strain (a) and compressive strain (b). (c, d) The Sn-$s$ and $p$ orbital weights in band-1 and 2 as a function of $c/a$. The character inversion occurs when $c/a$ is  below  0.93 and above 1.07. The inversion is accompanied with change in the topological invariant numbers (Z$_2$: $\nu_0$;$\nu_1$,$\nu_2$,$\nu_3$) leading to NI-TI phase transition. (e) Schematic illustration of formation of negative bandgap and subsequently the band inversion. } 
\label{fig2}
\end{figure}

\begin{figure*}
\centering
\includegraphics[angle=-0.0,origin=c,height=8.0cm,width=17.5cm]{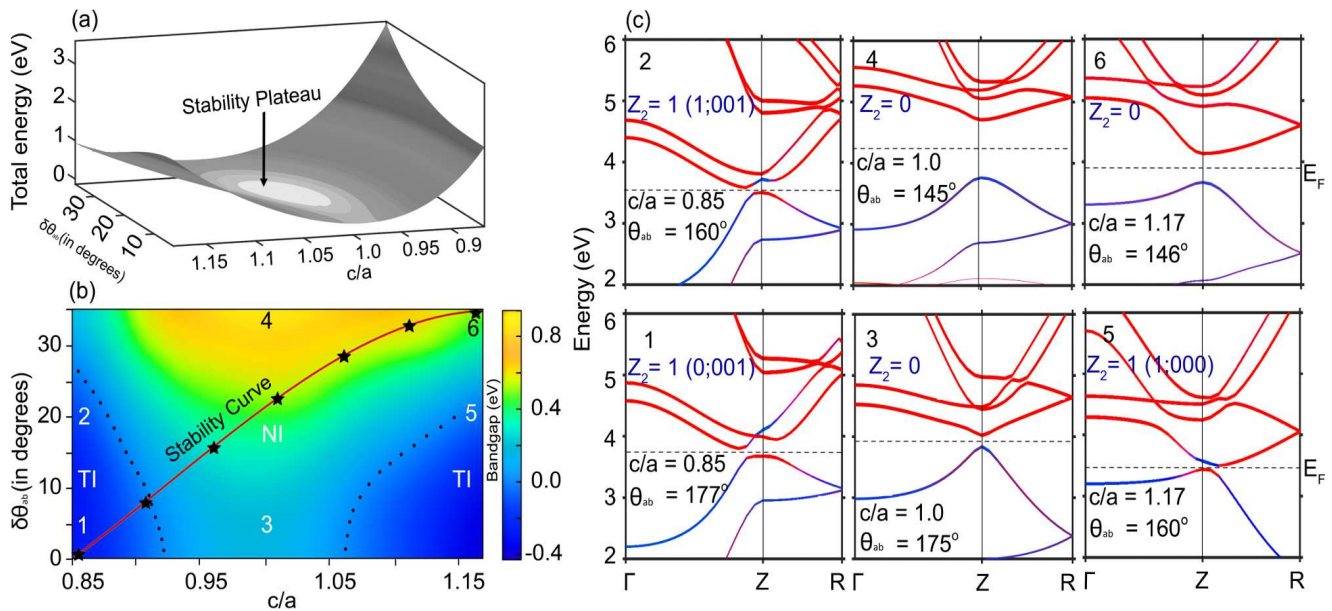}
\caption{(a) The variation of total energy (relative to ground state structure) and (b) bandgap as a function of $c/a$ and $\delta \theta_{ab}$ for $\beta$-CsSnI$_3$.  In this paper, we define $c/a$ as the ratio between the nearest Sn-Sn bond length - which is also the lattice parameter in the $\alpha$-phase - along $\hat{z}$ and along $\hat{x}$. The  color-bar indicates the magnitude of bandgap measured at the TRIM Z. The solid red curve in (b) represents the minimum energy curve or the stability curve with an optimized value of $\delta \theta_{ab}$.  The black dotted lines separate the NI from the TI phase.  The total energy and the bandgap maps are obtained through interpolation out of DFT calculated, suitably chosen 42 and 80 odd points respectively} in the $c/a$-$\delta \theta_{ab}$ plane. (c) Orbital resolved DFT band structure of some of the representative points (1 to 6) indicated in (b). As expected, the band inversion occurs for the points with negative bandgap. 
\label{fig3}
\end{figure*}

The nature of lattice distortion due to applied epitaxial strain depends on the ground state structure. In the $\alpha$-phase, the strain brings change to the translational symmetry as only  the $c/a$ ratio changes without tilting and rotating the octahedra. In  Fig. \ref{fig2}(a-b), we have shown the band structure of $\alpha$-CsSnI$_3$ for $c/a$ = 0.9, representing the tensile strain, and for $c/a$ = 1.1, representing the compressive strain. We see that at the TRIM $R$, the band character reverses with respect to the ground state structure (see Fig. \ref{fig1}) irrespective of the strain condition. Now, the Sn-p characters construct the VBM and the Sn-s character constructs the CBM. This band inversion brings a parity transformation to create a topologically invariant Dirac state at the surface as will be discussed in section-IV-C. To provide a quantitative measure of band inversion, in  Fig. \ref{fig2}(c-d) we have plotted the Sn-s and -p orbital weight at $R$ for band-1 - constituting the VBM - band-2 -constituting CBM - as a function of $c/a$. It shows that there exist two critical values of $c/a$ one each for compressive and tensile strain. Beyond these critical values, the band inversion occurs to induce the TI phase. This can be explained from the fact that with strain, the bond lengths either in the xy plane (compressive) or along the $z$-axis decreases. With the decrease in the bond length, there is substantial increase in the hopping interaction strength ($t$) and hence the bandwidth, particularly for band-1. As a consequence, the gap between band-2 and band-1 decreases. At a certain critical values of strain, the gap becomes negative leading to inversion (see schematic Fig. \ref{fig2}e). A quantitative measure of $t$s as a function strain is illustrated in Fig. \ref{fig7} and will be discussed in the context of tight-binding model. In the context of uniform hydrostatic pressure, the $t$ increases only with compression and hence there is one critical point for NI-TI phase transition \cite{Ravi, strain3}. 

In the equilibrium configuration, $\beta$-phase presents a reduced symmetry structure compared to the $\alpha$-phase as the linear Sn-I-Sn octahedral axis creates a bond angle ($\theta_{ab}$)  of 162$^\circ$ - deviating from the ideal $\alpha$-phase value of 180$^\circ$ in the $ab$-plane. Therefore, we have two degrees of freedom, $c/a$, and $\delta\theta_{ab}$ ( = 180$^\circ$ - $\theta_{ab}$) that can be exploited externally to modify the band structure. Fig. \ref{fig3}(a) maps the relative change in the total energy of CsSnI$_3$ as a function of $c/a$ and $\delta\theta_{ab}$. Also, it indicates the value of $\delta\theta_{ab}$ corresponding to the local energy minimum for a given value of $c/a$ (see solid line, Fig. \ref{fig3}(b).  
The figure establishes the following three aspects. Firstly, with tensile strain,  $\theta_{ab}$ approaches 180$^{\circ}$ and thereby the structure approaches the $\alpha$-phase. With compressive strain, the octahedra rotate with each other to distort the lattice even more. The distortion enhances the bandgap as can be observed from Fig. \ref{fig3}(b) where the E$_g$ is mapped in the space spanned by $c/a$ and $\delta\theta_{ab}$. 

Secondly, near the unstrained equilibrium configuration ($c/a$ = 1, $\delta\theta_{ab}$ = 18$^{\circ}$), a minor change in $\delta\theta_{ab}$ change the bandgap significantly though the total energy remains almost flat. We find that by varying the $\delta\theta_{ab}$ from 0 to 35$^{\circ}$, the total energy increases by a maximum of 400 meV. However, the bandgap can be increased approximately by 0.7 eV. Such a large scale tuning of bandgap, with relatively less cost in total energy, which probably can be supplied to the system by thermal means, has its significance in optoelectronic applications \cite{eg_eff}. Thirdly, with tensile strain negative bandgap emerges (see Fig. \ref{fig3}(b). In the negative gap region, the band inversion occurs which lead to the formation of topological insulating phase.  To further reconfirm and illustrate the role of the bandgap on the band inversion, we have identified six points  in Fig. \ref{fig3}(b) in the space spanned by $c/a$ and $\delta\theta$ and plotted the respective band structures in Fig. \ref{fig3}(c).  The points lie within the area enclosed by the dotted black line show the inversion of the $s$ and $p$ orbital characters between the VBM and CBM at $Z$. The other points exhibit the normal band insulator behavior.  Interestingly, the stability curve passes through the negative to positive bandgap region as the strain changes its character from compressive to expansive. Hence, unlikely the $\alpha$-phase, the $\beta$-phase can establish the TI phase only through tensile epitaxial strain.

\begin{figure}[h]
\centering
\includegraphics[angle=-0.0,origin=c,height=9cm,width=9cm]{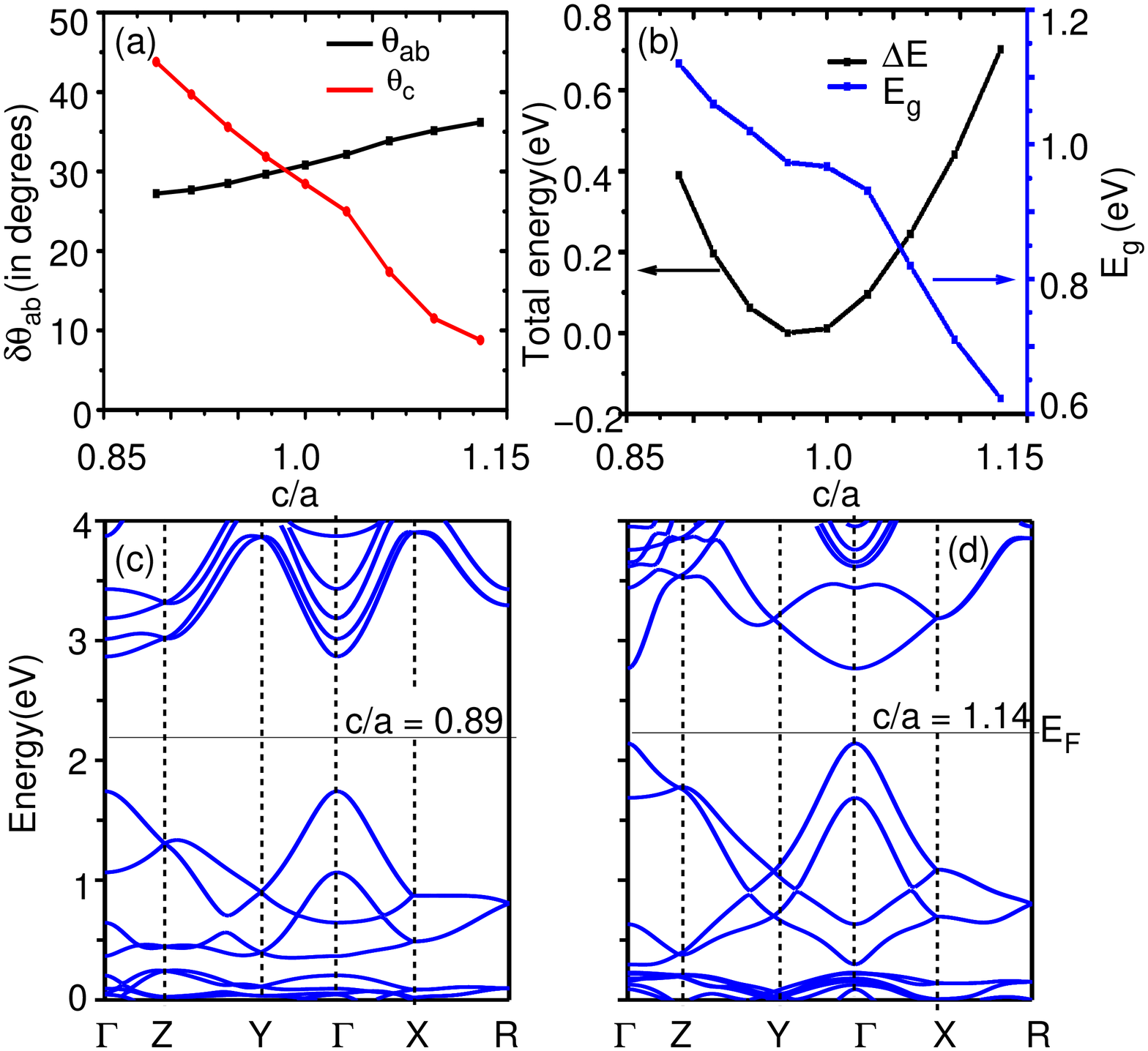}

\caption{(a) Change in the tilting and rotational angles,  $\delta\theta_{ab}$ and $\delta\theta_c$, and (b) the variation in total energy and bandgap as a function of $c/a$ for $\gamma$-CsSnI$_3$. (c, d) The representative band structures demonstrating the effect of tensile and compressive strain on the band topology.} 
\label{fig4}
\end{figure}

 As mentioned in section-II, the $\gamma$-phase represents the lowest symmetry structure among the three phases, and it is characterized by both in-plane ($\theta_{ab}$) and out of plane ($\theta_c$) octahedral rotations.
 Fig. \ref{fig4} summarizes the effect of strain on the stability and electronic structure of CsSnI$_3$ in this phase.  The optimized values of $\theta_{ab}$ and $\theta_{c}$ as a function of $c/a$ are shown in the  Fig. \ref{fig4}(a-b). The tensile  strain distorts the lattice more as both the angles deviate considerably from 180$^{\circ}$. However, the energetically expensive compressive strain brings the lattice close to the $\beta$-phase with $\theta_c$ approaching 180$^{\circ}$. The estimation of bandgap, plotted in  solid blue line, shows that larger the octahedral rotation and tilting, larger the bandgap. One can minimize the bandgap in this phase by compression.  However, we find that it is far from being negative even after 15$\%$ compression (bandgap = 0.6 eV) which further implies that it is very difficult to induce the TI phase in $\gamma$-CsSnI$_3$. 
 
\section{Model Hamiltonian for the Bulk  Halide Perovskites}
The modulation of the bandgap and hence the topological phase, as revealed from DFT calculations, are found to be dependent on (I) second neighbor Sn-Sn interactions ($t$) \cite{Ravi}, (II) the octahedral rotation/tilting angles $\theta_{ab}$, $\theta_c$, and (III) the $c/a$ ratio induced through strain. Therefore, formulation of an empirical relationship between E$_g$ and the influencing factors ($\epsilon$, $t$, $\theta$, and $c/a$) is desirable. Such a relationship can serve as the basis to establish a predictive design principle to tune the bandgap for specific applications in the area of optoelectronics and topology. In this regard, while the all-electron DFT calculations are best suitable, being computationally expensive, only a discrete set of such calculations can be carried out. To fill the void, the minimal basis set based tight-binding calculations, following the Slater-Koster formalism \cite{slater}, are adopted in this work. The appropriate SK-TB Hamiltonian for the family of ABX$_3$, in the second quantization notation, is expressed as  
\begin{equation}\label{1}
H = \sum_{i,m}\epsilon_{im}c_{im}^\dag c_{im} + \sum_{\langle\langle ij \rangle\rangle;m,n}t_{im jn}(c_{im}^\dag c_{jn} + h.c) + \lambda\textbf{L}\cdot\textbf{S}.
\end{equation}
Here, {\it i }({\it j}) and  $m$ ($n$ )   are site  and the orbitals ($s$, $p_x$, $p_y$, and $p_z$) indices respectively. The parameters $\epsilon_{im}$ and $t_{im jn}$ respectively represent the on-site energy and the strength of hopping integrals. The spin-orbit coupling (SOC) is included in the third term of the Hamiltonian with $\lambda$ denoting the SOC strength. The SOC requires doubling of the Hilbert space in the spin-domain. 
Since the unitcell of $\alpha$, $\beta$ and $\gamma$ phases consists of one, two and four formula units, the order of the Hamiltonian matrix for these phases in the same sequence is 8, 16 and 32. Furthermore, in the present consideration, the Hamiltonian needs to obey the time reversal symmetry which can be achieved by ensuring the following relation \cite{Book}.

\small{
\begin{subequations} \label{2}
\begin{equation} 
\mathbf{H}
 =  \left( \begin{array}{cc}
    \mathbf{H_{\uparrow\uparrow}}& \mathbf{H_{\uparrow\downarrow}} \\
    \mathbf{H_{\downarrow\uparrow}^{\dagger}}&  \mathbf{H_{\downarrow\downarrow}}
    %(= H_{\uparrow\uparrow}^{\dagger})
\end{array}  \right); 
 H_{\downarrow\downarrow} = (H_{\uparrow\uparrow})^*, H_{\uparrow\downarrow}
 =(H_{\downarrow\uparrow})^{\dagger}.
\end{equation}
While H$_{\downarrow\downarrow}$ is the time reversal partner of H$_{\uparrow\uparrow}$, the off-diagonal component (H$_{\uparrow\downarrow}$) is position independent  and its conjugate transpose enable the hopping between the partners. In the expanded form, these sub-matrices are found to be
\begin{equation}
 H_{\uparrow\uparrow}
 =  \left( \begin{array}{cccc}  
 
    \epsilon_s+f_0 & 2it_{sp}^xS_x & 2it_{sp}^xS_y & 2it_{sp}^zS_z\\
    -2it_{sp}^xS_x & \epsilon_{p}^x+f_1 & -i\lambda &0\\ 
    -2it_{sp}^xS_y & i\lambda & \epsilon_{p}^x+f_2 & 0\\
    -2it_{sp}^zS_z &   0 & 0 & \epsilon_{p}^z+f_3 
\end{array}  \right),
\end{equation}
\begin{equation}
H_{\uparrow\downarrow}
=\left( \begin{array}{cccc}
    0&0&0&0\\
    0&0&0&\lambda\\ 
    0&0&0&-i\lambda\\
    0&\lambda&-i\lambda&0
\end{array}  \right).
\end{equation}
\end{subequations}

 The dispersion relations $f$ are obtained by using the Slater-Koster TB relations \cite{slater}. For the (un)strained $\alpha$-phase they are
\begin{eqnarray}\label{3}
f_0& = &2t_{ss}^xcos(k_xa)+2t_{ss}^xcos(k_ya)+2t_{ss}^zcos(k_zc), \nonumber \\
f_1& = &2t_{pp\sigma }^xcos(k_xa)+2t_{pp\pi }^xcos(k_ya)+2t_{pp\pi }^zcos(k_zc), \nonumber \\
f_2& = &2t_{pp\sigma }^xcos(k_ya)+2t_{pp\pi }^xcos(k_xa)+2t_{pp\pi }^zcos(k_zc), \nonumber\\
f_3& = &2t_{pp\sigma }^zcos(k_zc)+2t_{pp\pi}^xcos(k_xa)+2t_{pp\pi }^xcos(k_ya).
\end{eqnarray}
}
The $f$ for the $\beta$ and $\gamma$-phases are provided in the appendix-A. The value of onsite energy ($\epsilon$) and the hopping integrals $t$ are listed in Table II, are estimated by fitting the TB band structure with that of the DFT. The TB band structure of CsSnI$_3$ for all the three phases, demonstrates an excellent agreement with that of DFT as can be observed from the  Fig. \ref{fig5}(a-c). It may be noted that the onsite parameters $\epsilon_s$ and $\epsilon_p$ are not same as the atomic energy level which is typically the case when a full basis, comprising of Sn-\{s, p\} and I-p orbitals, is considered. Here, they are in  the band centers of the antibonding $s$ and $p$ bands emerging out of the Sn-I nearest neighbor covalent interactions as discussed earlier. Therefore, $\epsilon_{p}^x$ ( = $\epsilon_{p}^y$) and $\epsilon_{p}^z$ differ with strain and octahedral distortions. Also, the strength of $t$ not necessarily represent the real Sn-Sn second neighbor interactions as they are remapped in order to accommodate the dispersion effect emerging out of the aforementioned covalent interactions  \cite{sro,Ravi}.

\begin{table}[h]
\centering
\caption{ Interaction parameters ($\epsilon$ and $t$) for unstrained equilibrium configurations in the units of eV. The SOC strength ($\lambda$) is estimated to be 0.14 eV. }
\begin{tabular}{cccccccc}
\hline
\hline
Phase  &  Interaction path& $\epsilon_s$ & $\epsilon_p$ & $t_{ss}$ & $t_{sp}$& $t_{pp\sigma}$ & $t_{pp\pi}$ \\
 \hline
$\alpha$  & $\hat{x}$,$\hat{y}$,$\hat{z}$ & 2.22  & 6.06 &-0.23 &	0.49 &  0.858&  0.09\\\\
$\beta$  &$\hat{x}$,$\hat{y}$             & 1.91  & 5.76 &-0.21 &	0.42 &	0.77 &	0.08  \\
          &$\hat{z}$                      & 1.91  & 5.8  &-0.22 &	0.46 &	0.82 &	0.09 \\\\
$\gamma$  & $\hat{x}$,$\hat{y}$           & 2.3   & 6.12 &-0.2  &	0.4  &	0.72 &	0.08 \\
          &$\hat{z}$                      & 2.3   & 6.37 &-0.24 &	0.4  &	0.86 &	0.09 \\
\hline
\hline
\end{tabular}
\begin{tablenotes}
      \small
      \item Note: In the case of $\gamma$-phase,  the in-plane lattice parameters are nearly equal. Hence, the difference in the interaction strength along $\hat{x}$ and $\hat{y}$ is negligible. 
    \end{tablenotes}
\end{table}

\subsection{A Common TB Model for all the Three Phases} 
\begin{figure}[h]
\centering
\includegraphics[angle=-0.0,origin=c,height=6.5cm,width=9.0cm]{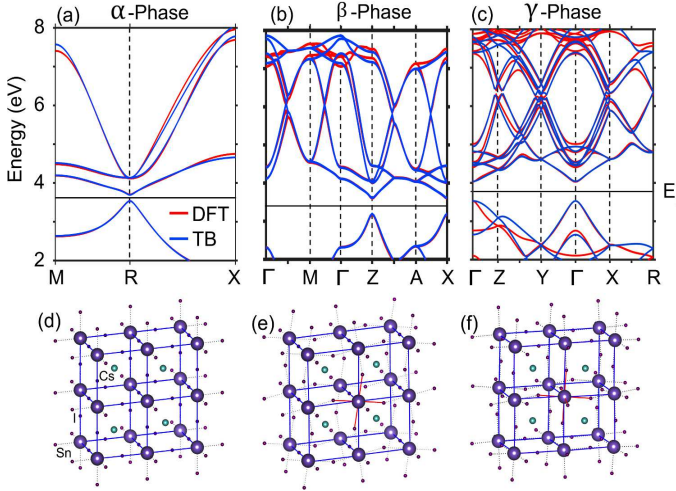}
\caption{ First row: DFT (red) and TB (blue) band structure of different phases of CsSnI$_3$. Second row: The cubic units of the $\alpha$-phase and the distorted-cubic units of lower symmetry $\beta$ and $\gamma$ phases. If I and Cs are ignored, the lattice of Sn alone (blue lines) in the $\beta$- and $\gamma$- phases can be represented as strained $\alpha$ phases.}
\label{fig5}
\end{figure}

\begin{figure}[h]
\centering
\includegraphics[angle=-0.0,origin=c,height=5.5cm,width=9.0cm]{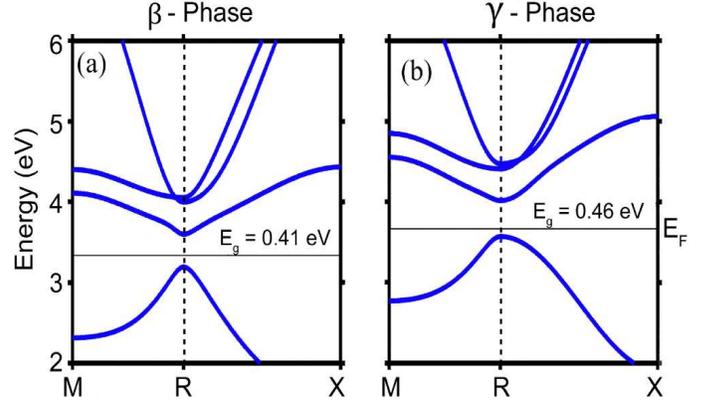}
\caption{ The band structure of unstrained $\beta$- and $\gamma$-CsSnI$_3$ as obtained with a minimal basis set based common tight-binding (CTB) model as discussed in the text. The TB parameters of Table II are used to calculate the band structure.}
\label{fig6}
\end{figure}

As we have now learnt that the nearest neighbor Sn-I covalent interaction can be implicitly included in the Hamiltonian through $\epsilon$ and $t$, the following  questions can be asked. Do we necessarily need the real crystal unit cell to construct a TB model in order to capture the essential electronic behavior in the vicinity of the Fermi surface? Can  we design a smaller unit cell which is sufficient to reduce the basis set of the TB Hamiltonian, so that the formation of a large intractable Hamiltonian matrix (e.g. the 32$\times$32 for the $\gamma$-phase) can be avoided? A closer inspection of the crystal structure of the three phases (see Fig. \ref{fig5} (d-f)) suggests that the cation Sn does not change its symmetry (Wyckoff) position due to change in the crystal phase. Hence, the lattice of Sn alone is same for all the phases, except the change in the length of the lattice vectors as can be noticed from the Fig. \ref{fig5}(d-f). The later is equivalent to application of anisotropic strain to the cubic lattice. Therefore, the TB model for the $\alpha$-phase can be accepted as a common TB model (CTB) for the family of halide perovskites. With this model, the phases will be distinguished from each other through the strength of the interaction parameters. Therefore, the dispersion relation of Eq. 3 remains the same for the CTB model except the fact that each of the $t$s are now explicitly function of c/a and $\theta$. The band structure for the $\beta$ and $\gamma$-phases, obtained with the CTB model, are shown in Fig. \ref{fig6} which shows that the bandgap is well reproduced. In fact, the TB band structure for the real unit cell of the $\beta$ and $\gamma$-phases can be reconstructed from the CTB band structure through band downfolding.

\subsection{Construction of a Descriptor Model}
\begin{figure}
\centering
\includegraphics[angle=-0.0,origin=c,height=11.0cm,width=8.0cm]{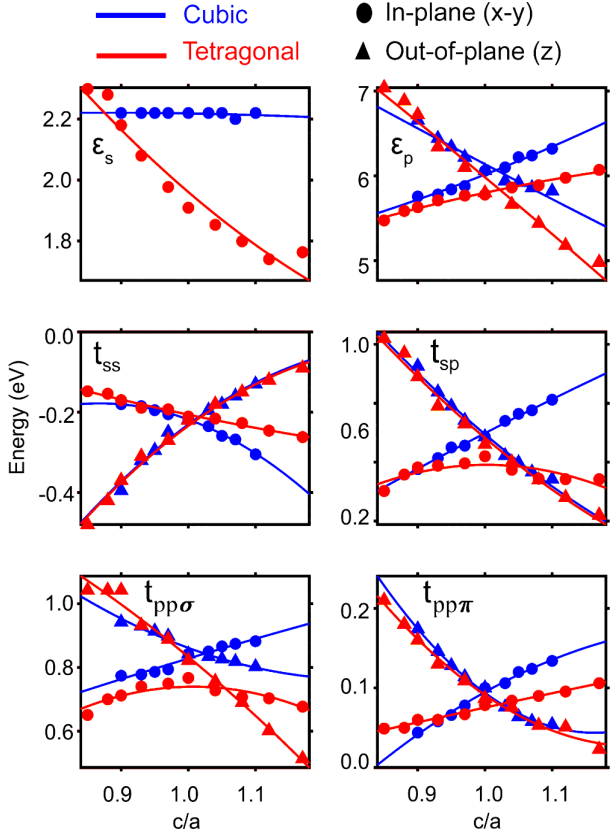}
\caption{ The variations of the in-plane as well as out-of-plane effective on-site energies and hopping interactions as a function of strain for $\alpha$ and $\beta$-phases of CsSnI$_3$. The quantitative deviations of the parameters between the $\alpha$ and $\beta$-phases are mainly attributed to the octahedral rotations of $\beta$-phase. As shown in Fig. \ref{fig3}(b), the octahedral angles change with strain to achieve a lower energy configuration.  
} 
\label{fig7}
\end{figure}

The structural optimization carried through DFT calculations in the previous sections have revealed that with strain, the bond lengths and bond  angles such as rotation and tilting (specific to the $\beta$ and $\gamma$-phases) change. As a result, the TB parameters change and consequently the band structure differs from the equilibrium configuration. Therefore, in order to gain insight into the band topology of the halide polymorphs, it is crucial to examine the variation of the TB parameters with strain. This has been carried out in this work for the case of $\alpha$ and $\beta$-phases and the results are illustrated in Fig. \ref{fig7}. The blue and red solid lines correspond to the strained $\alpha$ and $\beta$-phases respectively. In general, the interaction strengths increase with decrease in bond length and it is well reflected through the blue lines.  The deviation of the parameters of the $\beta$-phase from the $\alpha$-phase implies that not only the bond-length, but also the rotation/tilting angles have considerable influence on the band structure.

The effect of each of the parameters on the band structure in the vicinity of the Fermi level can be explained by calculating the eigenvalues at the TRIM $R$ by diagonalizing the CTB matrix of Eq. \eqref{2}. They form four  degenerate pairs as expressed below. 
\begin{eqnarray} \label{4}
E_{1, 2} & = & \epsilon_s - 4t_{ss}^x - 2t_{ss}^z, \nonumber \\
E_{3, 4} & = &  \frac{(\epsilon_{p}^x+\epsilon_{p}^z-\lambda)}{2}-2t_{pp\pi}^x-\zeta- \frac{\sqrt{8\lambda^2+(\lambda+\chi)^2}}{2},\nonumber \\
E_{5, 6}& = &  \epsilon_{p}^x + \lambda - 2t_{pp\pi}^x - 2t_{pp\pi}^z - 2t_{pp\sigma}^x, \nonumber \\
E_{7, 8} & = &  \frac{(\epsilon_{p}^x+\epsilon_{p}^z-\lambda)}{2}-2t_{pp\pi}^x-\zeta + \frac{\sqrt{8\lambda^2+(\lambda+\chi)^2}}{2}.
\end{eqnarray}
Where,
\begin{eqnarray} 
    \chi & = & \epsilon_{p}^x-\epsilon_{p}^z+2(t_{pp\pi}^z + t_{pp\sigma}^x - t_{pp\sigma}^z- t_{pp\pi}^x), \nonumber \\
     \zeta & = & t_{pp\pi}^z + t_{pp\sigma}^x + t_{pp\sigma}^z + t_{pp\pi}^x. \nonumber
\end{eqnarray}

The superscripts $x$ and $z$ refers to in-plane and out-of-plane interactions. The pair E$_{1,2}$ form the VBM and the pair E$_{3, 4}$ form the CBM. These two pairs determine the band topology as E$_{5,6}$ and E$_{7,8}$ lies above E$_{3, 4}$. The pre-factor to the TB parameters of E$_{1,2}$ and E$_{3,4}$ can be defined as topology-influencers (TIF), and they are listed in Table III.  
Fig. \ref{fig7}  and Table III together can be referred to provide a physical interpretation for TIF. For example, for change of 0.3 in $c/a$ ratio, the t$_{ss}^x$ changes by $\sim$ 0.1 eV. However, it shifts the VBM at $R$ by 0.4 eV. On the other hand,  though for the same change in $c/a$  t$_{sp}^x$ changes by 0.4 eV, it does not disturb the VBM and CBM at all.  
\begin{table}[h]
\centering
\caption{ The topological influencers  for different interaction parameters. }
\begin{tabular}{ccccccccc}
\hline
\hline
$\epsilon_s$ & $\epsilon_p^x$ & $\epsilon_p^z$ & $t_{ss}^x$ & $t_{ss}^z$& $t_{pp\sigma}^x$ & $t_{pp\sigma}^z$ & $t_{pp\pi}^x$ & $t_{pp\pi}^z$ \\
 \hline
1& $\frac{1}{2}$& $\frac{1}{2}$& 4 &2 &$1$ &$1$& $3$ & $1$ \\
\hline
\hline
\end{tabular}
\end{table}

After defining the influencers, which are independent of strain or any other external effect, it is crucial to describe the variation of the $\epsilon$ and $t$ as a function of strain so as to complete the predictive principle.  This has been achieved in the following two step processes. In the first step, the TB parameters are obtained for a discrete set of c/a or $\theta$ values by fitting with the realistic DFT results. As a proof,  In appendix-A, we have shown the TB fitted DFT band structure for a set of c/a and $\delta \theta_{ab}$ values. In the second step, the discrete set of interactions are fitted with polynomials of second order.  Higher order polynomial fitting is found to be inconsequential and hence ignored. The coefficients of the polynomial expansion can be regarded as a set of descriptor as they hold the information of band variation with strain. Altogether, the set of interactions [T], descriptors [D] and external controllers [EC] form a characteristic equation as given in  Eq. \eqref{5} and \eqref{6}. Also, specific to CsSnI$_3$ the value of descriptors are provided in these equations, and the fitted curves are shown with solid lines in Fig. \ref{fig7}.

\begin{figure}
\centering
\includegraphics[angle=-0.0,origin=c,height=5.0cm,width=9.0cm]{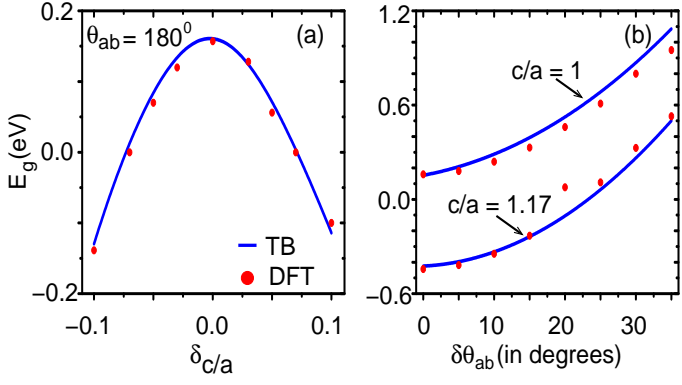}
\caption{ (a) Variation of the bandgap as a function of $c/a$ for constant $\theta_{ab}$ and (b) vice-versa  as estimated using the  descriptor method. The DFT bandgap values are denoted with red dots, which shows excellent agreement with the descriptor method.  } 
\label{fig8}
\end{figure}

\begin{widetext}
\begin{equation} \label{5}
  \left(\begin{array}{c}
      \epsilon_s\\
    \epsilon_{p}^x \\
           \epsilon_{p}^z \\
           t_{ss}^x \\
           t_{ss}^z\\
           t_{sp}^x \\
           t_{sp}^z \\
           t_{pp\sigma}^x\\
           t_{pp\sigma}^z\\
           t_{pp\pi}^x\\
           t_{pp\pi}^z
  \end{array}\right)
   =  \left(\begin{array}{ccc}
      \epsilon_{11} & \epsilon_{12} & \epsilon_{13} \\
          \epsilon_{21} & \epsilon_{22} & \epsilon_{23} \\
    \epsilon_{31} & \epsilon_{32} & \epsilon_{33} \\
     \eta_{11} & \eta_{12} & \eta_{13}\\
     \eta_{21} & \eta_{22} & \eta_{23}\\
     \eta_{31} & \eta_{32} & \eta_{33}\\
     \eta_{41} & \eta_{42} & \eta_{43}\\
          \eta_{51} & \eta_{52} & \eta_{53}\\
     \eta_{61} & \eta_{62} & \eta_{63}\\
     \eta_{71} & \eta_{72} & \eta_{73}\\
     \eta_{81} & \eta_{82} & \eta_{83}
  \end{array}\right)
  \left(\begin{array}{c}
  1\\
  \delta_{c/a} \\
  \delta_{c/a}^2 
  \end{array}\right)
  =  \left(\begin{array}{ccc}
  2.22 & 0 & 0\\
  6.01 &3.12&1.7 \\
    6.07 &4.2&16.9 \\
  -0.22 &-0.61&-2.33 \\
  -0.22 &1.23&-4.2 \\
  0.50 &1.18&-0.28 \\
  0.48 & -1.93 & 5.26\\
  0.83&0.63&-0.145 \\
  0.86 &-0.77&1.5 \\
  0.09&0.47&-0.64 \\
  0.09&0.61&1.89 
  \end{array}\right)
   \left(\begin{array}{c}
  1\\
  \delta_{c/a} \\
  \delta_{c/a}^2 
  \end{array}\right), \hspace{0.1cm }\mbox{for CsSnI$_3$.}
    \end{equation}
  \begin{equation}\label{6}
  \left(\begin{array}{c}
\epsilon_{s} \\
\epsilon_{p}^x \\
\epsilon_{p}^z \\
t_{ss}^x \\
t_{sp}^x \\
t_{pp\sigma}^x\\
t_{pp\pi}^x\\
\end{array}\right)
=
\left(\begin{array}{ccc}
\xi_{11} & \xi_{12} & \xi_{13} \\
\xi_{21} & \xi_{22} & \xi_{23} \\
\xi_{31} & \xi_{32} & \xi_{33} \\
\mu_{11} & \mu_{12} & \mu_{13}\\
\mu_{21} & \mu_{22} & \mu_{23}\\
\mu_{31} & \mu_{32} & \mu_{33}\\
\mu_{41} & \mu_{42} & \mu_{43}
\end{array}\right)
\left(\begin{array}{c}
1\\
\delta \theta \\
\delta \theta ^2 
\end{array}\right)
=
  \left(\begin{array}{ccc}
  2.19 & 0.57 & -0.402\\
6.0 & 0.14 & -1.11\\
  6.0 & 0.59 & 0.30\\
  -0.22 & 0.07 & 0.29\\
  0.50 & -0.04 & -0.86\\
  0.83 & 0.09 & -0.57\\
  0.09 & 0.04 & -0.04
  \end{array}\right)
  \left( \begin{array}{c}
  1\\
  \delta \theta \\
  \delta \theta^2 
  \end{array} \right), \hspace{0.1cm }\mbox{for CsSnI$_3$.}
\end{equation}

Note: Here, $\delta \theta$ is in terms of radian.  
\end{widetext}

Using Eq. \eqref{4} and the descriptor matrices of Eq. \eqref{5} and \eqref{6}, we can estimate the $E_g$ (= $E_{3,4} - E_{1,2}$) as a function of  $\delta c/a$ and/or $\delta \theta$ and few selected plots are shown in Fig. \ref{fig8}. We found that the bandgap obtained from above equations for constant $\delta \theta$ and $c/a$ shows a substantial agreement with DFT bandgap (red dots). These formulations assist in analyzing the bandgap as a function of strain. From the literature, it is evident that most of the halide perovskites  follow similar band structure \cite{Lamb_1,Lamb_2} and thus, the descriptor model universalizes the $E_g$ variation under different types of strain.
As the TB parameters are estimated by fitting the TB band structure with that of the DFT and it is expected that the error associated with the XC functionals will implicitly affect the accuracy of the parameters. To examine if that really is the case for the halides, in appendix B, we compared the TB parameters obtained within GGA+mBJ and HSE06. We find that except for a constant shift of the onsite energies, the variation of the TB parameters with strain remained almost the same. Therefore, the descriptors are expected to be independent of XC and hence have universal appeal.

\subsection{Surface Band Structure}
The bulk electronic structure study has enabled us to identify the possible  trivial and non-trivial phases and phase transition among them in CsSnI$_3$ polymorphs. However, the practical observation of these phases comes from the calculation of surface band structure. In this section, we develop  a Slater-Koster based TB Hamiltonian for slabs grown along (001) to obtain the surface band structure. Here, the slab consists of alternative stacking of CsI and SnI$_2$ layers. However, as discussed in section-II, there is no contribution of Cs and I orbitals at the Fermi level, the model is restricted to Sn-\{s, p\} orbital basis. The appropriate Hamiltonian for the slab consisting of $n$ unit cells (001) is 

\begin{equation}
  H_{slab} =\left (\begin{array}{cc}  
H_{\uparrow \uparrow}& H_{\uparrow \downarrow} \\
H_{\downarrow \uparrow}&  H_{\downarrow \downarrow}
  \end{array} \right).
\end{equation}
Where, 
\begin{equation}
    H_{\uparrow \uparrow} = 
    \left ( \begin{array}{cccccc} 
    H_{11} & H_{12} & 0 & 0 & 0&\dots \\
    H_{21} & H_{22} & H_{23} & 0 &0& \dots \\
    0 & H_{32} & H_{33} & H_{34} & 0&\dots \\
    \vdots & &\ddots&\ddots & \ddots \\
    \dots  & 0 &0&H_{n-1n-2}& H_{n-1n-1} & H_{n-1n}\\
    \dots & 0 & 0 &0& H_{nn-1} & H_{nn}
    \end{array} \right)
\end{equation}
Here, $H_{jj}$ is the Hamiltonian for $j^{th}$ layer of the slab,  and it is given by,
\begin{equation}
 H_{jj}= 
  \left( \begin{array}{cccc}  
    \epsilon_s+f_0 & 2it_{sp}^xS_x & 2it_{sp}^xS_y & 0\\
    -2it_{sp}^xS_x & \epsilon_{p}^x+f_1 & -i\lambda &0\\ 
    -2it_{sp}^xS_y & i\lambda & \epsilon_{p}^x+f_2 & 0\\
    0 &   0 & 0 & \epsilon_{p}^z+f_3 
\end{array}  \right)
\end{equation}
Here,
\begin{eqnarray}\label{3}
f_0& = &2t_{ss}^xcos(k_xa)+2t_{ss}^xcos(k_ya), \nonumber \\
f_1& = &2t_{pp\sigma }^xcos(k_xa)+2t_{pp\pi }^xcos(k_ya), \nonumber \\
f_2& = &2t_{pp\sigma }^xcos(k_ya)+2t_{pp\pi }^xcos(k_xa), \nonumber\\
f_3& = &2t_{pp\pi}^xcos(k_xa)+2t_{pp\pi }^xcos(k_ya).
\end{eqnarray}

The block H$_{jj-1}$ describe the interaction between layer j and j-1.  
\begin{equation}
H_{j-1j} = (H_{jj-1})^{T} = 
\left( \begin{array}{cccc}
        t_{ss}^z&0&0&t_{sp}^z\\
       0&t_{pp\pi}^z&0&0\\
    0&0&t_{pp\pi}^z&0\\  
    -t_{sp}^z&0&0&t_{pp\sigma}^z 
\end{array}  \right)
\end{equation}
The off diagonal block emerging from the SOC is of the form
\begin{equation}
    H_{\uparrow \downarrow}= 
    \left (\begin{array}{cccccc} 
    G_{11} & 0 & 0 & 0 & 0&\dots \\
    0 & G_{22} & 0 & 0 &0& \dots \\
    0 & 0& G_{33} & 0 & 0&\dots \\
    \vdots & &\ddots&\ddots & \ddots \\
    \dots  & 0 &0& 0&G_{n-1n-1} &0\\
    \dots & 0 & 0 &0& 0 & G_{nn}
    \end{array} \right )
\end{equation}
where, 
\begin{equation}
G_{jj}=\left( \begin{array}{cccc}
        0&0&0&0\\
    0&0&0&\lambda\\ 
    0&0&0&-i\lambda\\
    0&\lambda&-i\lambda&0
\end{array}  \right). 
\end{equation}
As mentioned in Eq. (\ref{2}), $H_{\downarrow \downarrow} = (H_{\uparrow \uparrow})^*$ and $H_{\downarrow \uparrow}=H_{\uparrow \downarrow}^\dagger$, which assures the  time reversal symmetry of the Hamiltonian matrix \cite{Book}.

\begin{figure}
\centering
\includegraphics[angle=-0.0,origin=c,height=9.7cm,width=8.5cm]{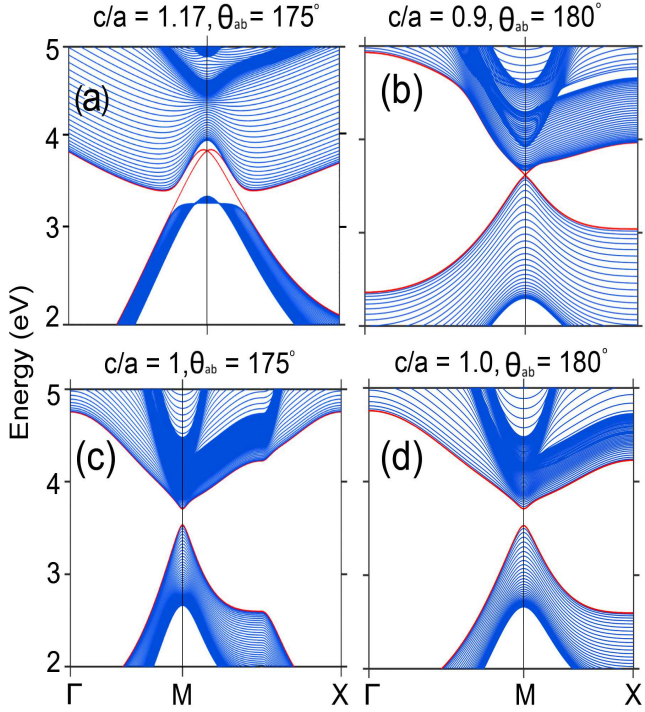}
\caption{Surface band structure of slabs constructed out of different bulk configurations of CsSnI$_3$ obtained from S-K based TB model. Existence of Dirac kind of bands in surface band structure at the bulk bandgap confirms the non-trivial insulating phase. } 
\label{fig9}
\end{figure}

 Considering a slab of 30 layers thick, in Fig. \ref{fig9}, we have shown the surface band structure of $\alpha$- and $\beta$- CsSnI$_3$ for different strain conditions. Bands with deep blue color correspond to bulk structure and bands with red color are from the surface. The figure justifies the prediction that,  s-p band inversion arising from negative bulk bandgap  yields TI phase. For example, in the upper row of the figure the strain condition creates negative E$_g$ and hence invariant surface Dirac states, which is a must for the TI phase, is formed. However, for the lower row, the strain condition is such that it creates positive E$_g$. Therefore, no invariant Dirac states are formed.  We find that the SK-TB formalism is appropriate to examine the surface phenomena and  it can be considered as an alternate tool besides the widely used $\textbf{\textit{k.p}}$ theory and Wannier formalism. 
 
\section{Summary and Outlook}
In summary, we have taken CsSnI$_3$ as a prototype for examining the electronic structure of  sp-element based halide perovskites which are promising for optoelectronic applications and possess the right crystal and orbital symmetries to demonstrate non-trivial quantum states. However, bandgap engineering is crucial to realize either of these two aspects.  In this work, we have adopted a dual approach, ab-initio DFT studies and design of parametric tight-binding Hamiltonian based on the framework of LCAO formalism and came up with predictive principles that are fundamental to carry out the bandgap engineering via strain in the halide perovskite family. Some of the crucial conclusions are as follows. (I) In the cubic phase, one can apply both compressive and tensile strains to reduce the bandgap from the equilibrium bulk value and eventually beyond a critical value the system exhibits negative bandgap and therefore the non-trivial topological insulator phase. (II) In the tetragonal phase, while tensile strain reduces the bandgap and thereby there is a possibility of normal insulator to topological insulator phase transition,  compressive strain enhances it and hence can be used as a controllable factor for optoelectronic applications. (III) Interestingly, the halide structure establishes stability plateau in the configuration space spanned by strain and octahedral rotations in the crystal. In this plateau, while the total energy of the system remains almost flat, the variation range of the bandgap is close to 1 eV.  (IV) The orthorhombic phase enhances the bandgap with tensile strain and reduces it with the compressive strain. Also, we find that the compressive strain tries to improve the symmetry as system gradually enters to the tetragonal phase, but at a higher energy cost. (V) Our model study led to conceptualization of topological influencers (TIF) which basically provide a quantitative measure of the ability of each electron hopping interaction to alter the band structure at the Fermi surface/chemical potential and hence the band topology of the system. These influencers are independent of chemical composition of the compound and hence have universal appeal. (VI) We realize that a basis set consisting of only four orbitals (e.g. Sn-\{s, p\}) is sufficient to include in the tight-binding (TB) Hamiltonian so that the electronic structure of each of the structural phases can be explained. (VII) The TB studies further led to the design of a descriptor model which describes the effect of structural distortion due to strain on the interaction parameters.

\begin{figure}
\centering
\includegraphics[angle=-0.0,origin=c,height=9.0cm,width=8.5cm]{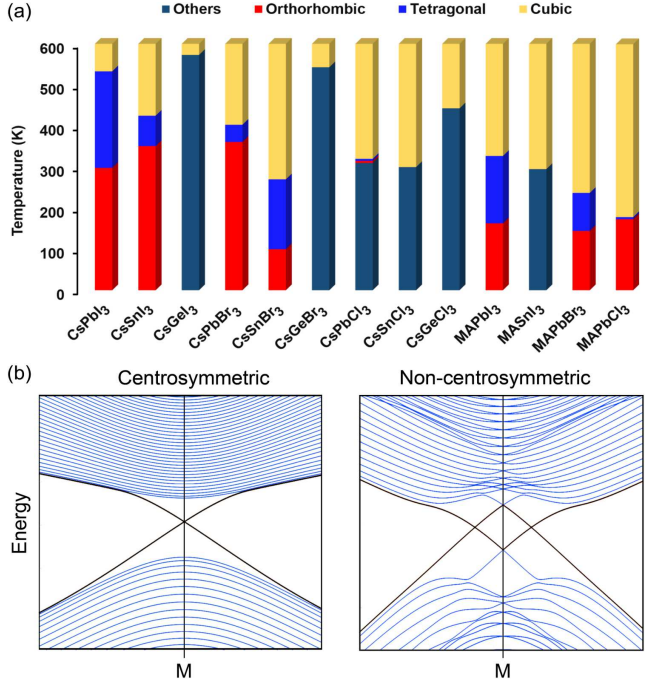}
\caption{(a) Bar chart representing the various crystal phases of inorganic and organic halide perovskites. The data is collected from different literature works \cite{1,2,3,struct2,6,7,10,11,13,struct1,Lamb_1,Lamb_2}. (b) Surface states of centrosymmetric and non-centrosymmetric systems. The later includes Rashba coupling that arises due to the inversion symmetry breaking electric field that exists in non-centrosymmetric systems. } 
\label{fig10}
\end{figure}

The value of the critical temperatures at which the structural phase transition occur, varies with the chemical composition of the perovskite. As can be seen from Fig. \ref{fig10}a, while some of the compounds have cubic phase at the room temperature, some others have tetragonal or orthorhombic phases. Therefore, the conclusions (I - IV) listed above can act as a fundamental guideline to room temperature engineering of the bandgap, with respect to the equilibrium bulk value, of a given material via strain. Accordingly, the application perspective can be decided.

Furthermore, centrosymmetricity of the compound plays a vital role in the electronic properties of the halides. There are first principles calculations which suggests that the organic halides like MAPbI$_3$ become non-centrosymmetric\cite{rashba_soc}. However, this is debated in the experimental studies. While some of the studies report about the noncentrosymmetric phase \cite{PNAS_non_centro}, some others have predicted the stability of centrosymmetric phase\cite{DDsarma}. If the halide perovskite is non-centrosymmetric, an internal electric field is created, leading to the removal of momentum degeneracy through Rashba coupling. Therefore, such a coupling should be considered while designing a model Hamiltonian, with the objective of explaining the band phenomena of all members of halide perovskites (inorganic, organic and hybrid). We find that the TB model in present paper addresses this aspect. When the Rashba coupling is included in the designed TB Hamiltonian (i.e.   $ H_{\mbox{centro}} + \alpha_R(\hat{k}\times\hat{z})\cdot \Vec{\sigma}$), it provides the appropriate band structure. For example, in Fig. \ref{fig10}b, we show that the invariant surface Dirac state, exhibited by a centrosymmetric perovskite, yield  two interpenetrating Dirac states with the inclusion of Rashba coupling. This is due to the breakdown of Kramers degeneracy in noncentrosymmetric structure, and it is a well established fact\cite{Book}. Therefore, from the topological perspective, the present work provides a universal TB Hamiltonian and also proposes a new set of numbers called TIF to engineer the topology of a given halide perovskites.

The critical strains at which the band inversion occurs for the $\alpha$ and $\beta$ phases  are proportional to the bandgap of the unstrained compound. The present study involves GGA XC potential with mBJ correction as well as the hybrid XC functional (see appendix-B) which are known to underestimate the bandgap of the halides and therefore, the measured critical strain values to induce the NI-TI phase transitions, as shown in Fig. \ref{fig2} and \ref{fig3}, are also expected to be underestimated. The approximated value of underestimation can be extrapolated from one of
our earlier studies on the effect of hyrdrostatic pressure on the electronic structure of $\alpha$-CsSnX$_3$ (X = I, Br, Cl)\cite{Ravi}. Here, the equilibrium the GGA-mBJ obtained bandgaps are 0.16, 0.42, 1.08 eV for I, Br, and Cl  respectively. We find that that the NI-TI phase transition occurs at critical compression ($a/a_0$) of about 0.97 (I), 0.94 (Br), and 0.91 (Cl).  Since GW calculations, which are experimentally more accepted, estimate the bandgap of $\alpha$-CsSnI$_3$ as $\sim$1 eV \cite{Lamb_1,Lamb_2}, one may borrow the critical compression value of $\alpha$-CsSnCl$_3$ as an approximated experimental value for the iodide compound. If we convert that in terms of Sn-Sn bond lengths, we will require $\sim$ 8\% and $\sim$6\%  tensile and compressive strains, to experimentally induce the phase transition in $\alpha$-CsSnI$_3$. It may be noted that the inorganic halide perovskites are known to be soft materials \cite{soft} and  a recent experimental study by Chen et al. \cite{expt_strain}  shows the  CsSnI$_3$ can withstand a 13\% of strain when grown on a mica substrate.

\section*{\small Acknowledgment} This work is supported by Department of Science and Technology, India through Grant No.EMR/2016/003791.
\newpage
\appendix

\section{Tight binding model for lower symmetry polymorphs of CsSnI$_3$:}

The tight binding Hamiltonian  discussed in the main text (Eq. \eqref{1} and Eq. \eqref{2}) for $\alpha$-phase is extendable to lower symmetry $\beta$ and $\gamma$-phases with appropriate number of basis. Here, we elaborate the TB Hamiltonian for $\beta$ and $\gamma$-phases to obtain the TB band structure.

\subsection{$\beta$-phase:}

\begin{figure}[h]
\centering
\includegraphics[angle=-0.0,origin=c,height=5.0cm,width=5.0cm]{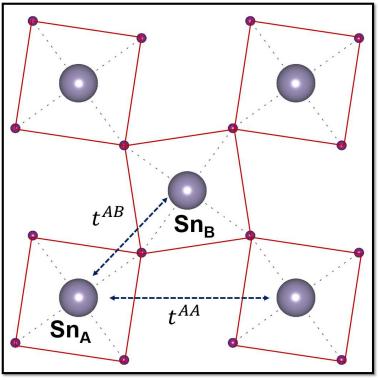}
\caption{The schematic representation of different hopping interactions in $ab$-plane of $\beta$-CsSnI$_3$. Here, the $t^{AA}$ hopping interaction are weak as compared to $t^{AB}$, as they do not involve Iodine mediation.  }
\label{fig1s}
\end{figure}

\begin{figure}
\centering
\includegraphics[angle=-0.0,origin=c,height=7.5cm,width=8.5cm]{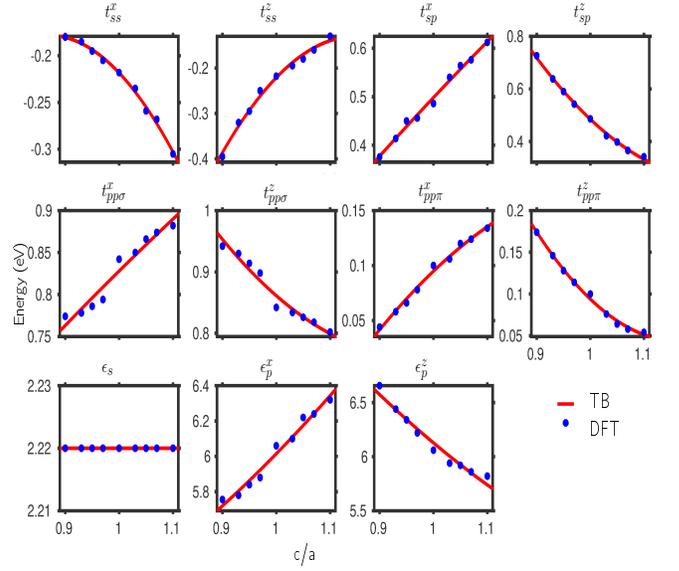}
\caption{Variation of different interaction parameters of $\beta$-CsSnI$_3$ as a function $c/a$ for a constant $\theta_{ab}$ ($= 162^{\circ}$).  }
\label{fig2s}
\end{figure}

\begin{figure}[h]
\centering
\includegraphics[angle=-0.0,origin=c,height=8.5cm,width=8.0cm]{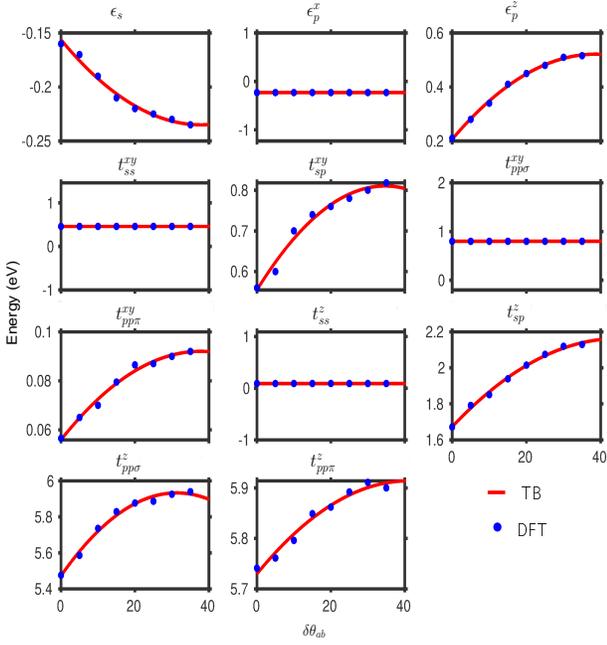}
\caption{Variation of different interaction parameters of $\beta$-CsSnI$_3$ as a function $\theta_{ab}$ for constant $c/a$ ($= 1$).  }
\label{fig3s}
\end{figure}

\begin{figure}
\centering
\includegraphics[angle=-0.0,origin=c,height=6.2cm,width=8.5cm]{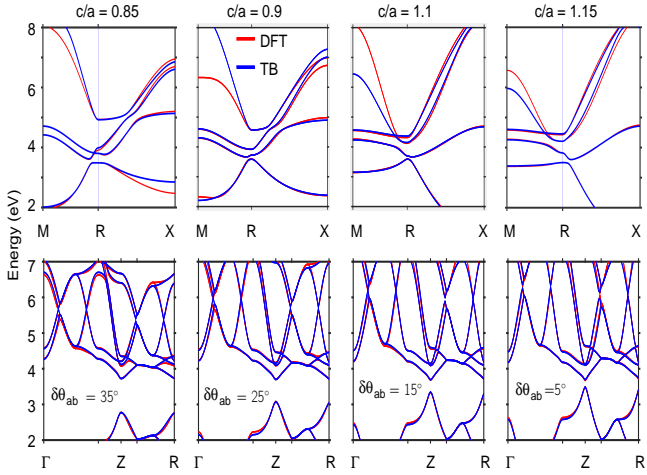}
\caption{TB fitted DFT band structure of $\alpha$($\beta$)-phase in first(second) row for different set of c/a and $\delta \theta_{ab}$ values.  }
\label{fig4s}
\end{figure}

 As discussed in section-II of the main text, the presence of octahedral rotation in $ab$-plane in $\beta$-phase creates two in-equivalent Sn atoms in the unit cell. We denote them as Sn$_A$ and Sn$_B$. Now the basis set include eight eigenstates, \textit{viz},  
$|s^{Sn_A}\rangle$, $|p^{Sn_A}_{x}\rangle$, $|p^{Sn_A}_{y}\rangle$, $|p^{Sn_A}_{z}\rangle$,$|s^{Sn_B}\rangle$, $|p^{Sn_B}_{x}\rangle$, $|p^{Sn_B}_{y}\rangle$, $|p^{Sn_B}_{z}\rangle$. The corresponding SOC included 16$\times$16 Hamiltonian matrix is given as,

\[
H_{TB}
 =  \left( \begin{array}{cc}
    H_{\uparrow\uparrow}& H_{\uparrow\downarrow} \\
    H_{\downarrow\uparrow}^{\dagger}&  H_{\downarrow\downarrow}
    %(= H_{\uparrow\uparrow}^{\dagger})
\end{array}  \right), \hspace{0.5cm}
H_{\uparrow \uparrow}
= (H_{\downarrow \downarrow})^*
 =  \left( \begin{array}{cc}
    H^{AA}& H^{AB} \\
    (H^{AB})^{\dagger}&  H^{BB}
    %(= H_{\uparrow\uparrow}^{\dagger})
\end{array}  \right)^{\uparrow\uparrow}, 
\]
and
\begin{equation}
\footnotesize
 H^{AA}_{{\uparrow\uparrow}} 
 =  \left( \begin{array}{cccc}
    \epsilon_s+f_0 & 2i(t_{sp}^x)^{AA}S_x & 2i(t_{sp}^x)^{AA}S_y & 2i(t_{sp}^z)^{AA}S_z\\
    -2i(t_{sp}^x)^{AA}S_x & \epsilon_{p}^x+f_1 & -i\lambda &0\\ 
    -2i(t_{sp}^x)^{AA}S_y & i\lambda & \epsilon_{p}^x+f_2 & 0\\
    -2i(t_{sp}^z)^{AA}S_z &   0 & 0 & \epsilon_{p}^z+f_3 
\end{array}  \right)
\end{equation}
\begin{equation}
 H^{AB}_{\uparrow\uparrow} 
 =  \left( \begin{array}{cccc}
    g_0 & g_1 & g_2 & 0\\
    g_1^* & g_4 & g_5& 0\\ 
    g_2^* & g_5^* & g_4 & 0\\
    0 &   0 & 0 & g_6 
\end{array}  \right)
\end{equation}

\begin{eqnarray}
f_0& = &2(t_{ss}^x)^{AA}(cos(k_xa)+cos(k_ya))+2(t_{ss}^z)^{AA}cos(k_zc) \nonumber \\
f_1& = &2(t_{pp\sigma}^x)^{AA}cos(k_xa)+2(t_{pp\pi}^x)^{AA}cos(k_ya)+2(t_{pp\pi }^z)^{AA}cos(k_zc) \nonumber \\
f_2& = &2(t_{pp\sigma}^x)^{AA}cos(k_ya)+2(t_{pp\pi}^x)^{AA}cos(k_xa)+2(t_{pp\pi }^z)^{AA}cos(k_zc)\nonumber\\
f_3& = &2(t_{pp\sigma}^z)^{AA}cos(k_zc)\nonumber\\
    g_0& = &4t_{ss}^{AB}cos(\frac{k_xa}{2})cos(\frac{k_ya}{2}) \nonumber\\
        g_1 & = & 2\sqrt{2}it_{sp}^{AB}sin(\frac{k_xa}{2})cos(\frac{k_ya}{2})\nonumber\\
    g_2& = &2\sqrt{2}it_{sp}^{AB}sin(\frac{k_ya}{2})cos(\frac{k_xa}{2})\nonumber\\
    g_4 & = & 2(t_{pp\sigma}^{AB}+t_{pp\pi}^{AB}) cos(\frac{k_xa}{2})cos(\frac{k_ya}{2})\nonumber\\
    g_5& = & 2(-t_{pp\sigma}^{AB}+t_{pp\pi}^{AB}) sin(\frac{k_xa}{2})sin(\frac{k_ya}{2})\nonumber\\
    g_6 & = & 4t_{pp\pi}^{AB} cos(\frac{k_xa}{2})cos(\frac{k_ya}{2})
    \end{eqnarray}
\begin{equation}
H_{\uparrow\downarrow}
=\left( \begin{array}{cccc}
  H_{\uparrow\downarrow}^{AA}&0\\
    0&H_{\uparrow\downarrow}^{BB}
\end{array}  \right).
\hspace{0.25cm}
H_{\uparrow\downarrow}^{AA} = H_{\uparrow\downarrow}^{BB}
=\left( \begin{array}{cccc}
    0&0&0&0\\
    0&0&0&\lambda\\ 
    0&0&0&-i\lambda\\
    0&\lambda&-i\lambda&0
\end{array}  \right).
\end{equation}
    Here, the $(t_i^{x/z})^{AA}$ interaction strengths are weak ($\sim 0.01$ eV) as they do not involve iodine  mediated hopping, we can safely neglect them. The significant parameters are listed in Table II of main text. In Fig. \ref{fig7} of the main text, for $\beta$-CsSnI$_3$, we have shown the variation of various interactions parameters as a function of $c/a$, where the deviation from $\alpha$-phase is due to existence of in-plane octahedral rotations. Here, in Fig. \ref{fig2s} and \ref{fig3s}, we present the effect of $c/a$ for constant $\theta_{ab}$ and vice versa. The validity of the TB parameters is shown in Fig. \ref{fig4s} by fitting the TB and DFT bands for representative values of $c/a$ and $\theta_{ab}$.

\subsection{Orthorombic phase:}
\begin{figure}
\centering
\includegraphics[angle=-0.0,origin=c,height=5.0cm,width=5.0cm]{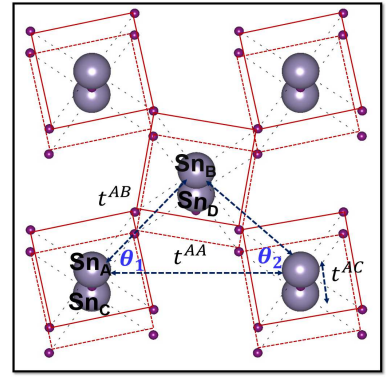}
\caption{The schematic representation of different hopping interactions in $ab$-plane and along $c$ direction of $\gamma$-CsSnI$_3$.   }
\label{fig5s}
\end{figure}
The $\gamma$-phase of CsSnI$_3$ is characterized by in plane ($ab$) rotation as well as out of plane (along $c$) octahedral tilting. This dual rotation creates four inequivalent Sn atoms. We denote them as Sn$_A$, Sn$_B$, Sn$_C$ and Sn$_D$. Therefore, the basis set augmented and include sixteen eigenstates and the corresponding SOC included 32$\times$ 32 Hamiltonian matrix is given as,

\[
H_{TB}
 =  \left( \begin{array}{cc}
    H_{\uparrow\uparrow}& H_{\uparrow\downarrow} \\
    H_{\downarrow\uparrow}^{\dagger}&  H_{\downarrow\downarrow}
    %(= H_{\uparrow\uparrow}^{\dagger})
\end{array}  \right),
\]
\[
H_{\uparrow \uparrow}
= H_{\downarrow \downarrow}^*
 =  \left( \begin{array}{cccc}
    H^{AA}& H^{AB} & H^{AC}& H^{AD}\\
    (H^{AB})^{\dagger} & H^{BB} & H^{BC}& H^{BD}\\
    (H^{AC})^{\dagger} & (H^{BC})^{\dagger} & H^{CC}& H^{CD}\\
    (H^{AD})^{\dagger} & (H^{BD})^{\dagger} & (H^{CD})^{\dagger} & H^{DD}\\
    %(= H_{\uparrow\uparrow}^{\dagger})
\end{array}  \right)^{\uparrow\uparrow}, \]
and
\begin{equation}
 H^{ii}_{\uparrow\uparrow} 
 =  \left( \begin{array}{cccc}
    \epsilon_s+f_0 & 2i(t_{sp}^x)^{AA}S_x & 2i(t_{sp}^x)^{AA}S_y & 0\\
    -2i(t_{sp}^x)^{AA}S_x & \epsilon_{p}^x+f_1 & -i\lambda &0\\ 
    -2i(t_{sp}^x)^{AA}S_y & i\lambda & \epsilon_{p}^x+f_2 & 0\\
    0 &   0 & 0 & \epsilon_{p}^z+f_3 
\end{array}  \right),
\end{equation}
Where, i=A, B, C and D.
\begin{equation}
 H^{AB}_{\uparrow\uparrow}= H^{CD}_{\uparrow\uparrow}
 =  \left( \begin{array}{cccc}
    g_0 & g_1 & g_2 & 0\\
    g_1^* & g_3 & g_5& 0\\ 
    g_2^* & g_5^* & g_4 & 0\\
    0 &   0 & 0 & g_6 
\end{array}  \right)
\end{equation}
\begin{equation}
 H^{AC}_{\uparrow\uparrow}= H^{BD}_{\uparrow\uparrow}
 =  \left( \begin{array}{cccc}
    f_4 & 0 & 0 & f_5\\
    0 & f_6 & 0 & 0\\ 
    0 & 0 & f_6 & 0\\
    f_5^* &   0 & 0 & f_7 
\end{array}  \right),
\end{equation}
\begin{equation}
 H^{AD}_{\uparrow\uparrow}= H^{BC}_{\uparrow\uparrow}
 =  \left( \begin{array}{cccc}
    0 & 0 & 0 & 0\\
    0 & 0 & 0 & 0\\ 
    0 & 0 & 0 & 0\\
    0 & 0 & 0 & 0 
\end{array}  \right),
\end{equation}

\begin{eqnarray}
f_0& = &2(t_{ss}^x)^{AA}cos(k_xa)+2(t_{ss}^y)^{AA}cos(k_yb) \nonumber \\
f_1& = &2(t_{pp\sigma }^x)^{AA}cos(k_xa)+2(t_{pp\pi }^y)^{AA}cos(k_yb) \nonumber \\
f_2& = &2(t_{pp\sigma}^y)^{AA}cos(k_yb)+2(t_{pp\pi}^x)^{AA}cos(k_xa) \nonumber\\
f_3& = &2(t_{pp\pi }^x)^{AA}cos(k_xa)+2(t_{pp\pi }^y)^{AA}cos(k_yb) \nonumber\\
f_4& = &2(t_{ss }^z)^{AC} cos{\frac{k_zc}{2}} \nonumber\\
f_5& = &2i(t_{sp }^z)^{AC} sin{\frac{k_zc}{2}} \nonumber\\
f_6& = &2(t_{pp\pi }^z)^{AC} cos{\frac{k_zc}{2}}\nonumber\\
f_7& = &2(t_{pp\sigma }^z)^{AC} cos{\frac{k_zc}{2}} \nonumber\\
\end{eqnarray}
\begin{eqnarray}
g_0& = &4t_{ss}^{AB}cos(\frac{k_xa}{2})cos(\frac{k_yb}{2}) \nonumber \\
g_1 & = & 4icos(\theta_1)t_{sp}^{AB}sin(\frac{k_xa}{2})cos(\frac{k_yb}{2})\nonumber\\
g_2& = & 4icos(\theta_{2})t_{sp}^{AB}sin(\frac{k_yb}{2})cos(\frac{k_xa}{2})\nonumber\\
g_3 & = & 4f({\theta_1}) cos(\frac{k_xa}{2})cos(\frac{k_yb}{2})\nonumber\\
g_4 & = & 4f(\theta_2) cos(\frac{k_xa}{2})cos(\frac{k_yb}{2})\nonumber\\
g_5 & = & 4f(\theta_{12}) sin(\frac{k_xa}{2})sin(\frac{k_yb}{2})\nonumber\\
g_6 & = & 4t_{pp\pi}^{AB} cos(\frac{k_xa}{2})cos(\frac{k_yb}{2})\nonumber\\
f(\theta_1)& = &(t_{pp\sigma }^{xy})^{AB}cos^2(\theta_1) + (t_{pp\pi }^{xy})^{AB}sin^2(\theta_1)\nonumber\\
f(\theta_2)& = &(t_{pp\sigma }^{xy})^{AB}cos^2(\theta_2) + (t_{pp\pi}^{xy})^{AB}sin^2(\theta_2)\nonumber\\
f(\theta_{12)}& = &((t_{pp\sigma }^{xy})^{AB}-(t_{pp\pi}^{xy})^{AB})cos(\theta_1)cos(\theta_2)\nonumber\\
\end{eqnarray}

Here, $\theta_1$ and $\theta_2$ arise due to slight difference in the lattice parameter $a$ and $b$.

\section{Effect of Exchange-Correlation functional on Bandgap and NI-TI transition}

\begin{figure}
\centering
\includegraphics[angle=-0.0,origin=c,height=5.5cm,width=5.5cm]{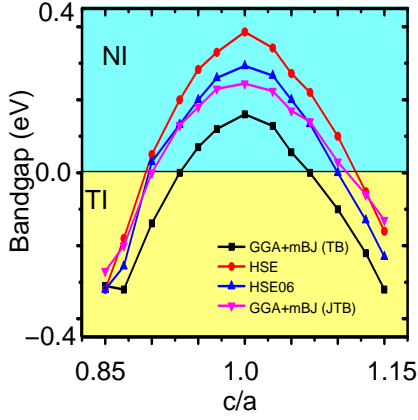}
\caption{ Variation of bandgap in $\alpha$-CsSnI$_3$ as function of biaxial strain for GGA+mBJ(TB),GGA+mBJ(JTB), HSE06 and HSE functionals.}
\label{fig6s}
\end{figure}

\begin{figure}
\centering
\includegraphics[angle=-0.0,origin=c,height=7.5cm,width=9.0cm]{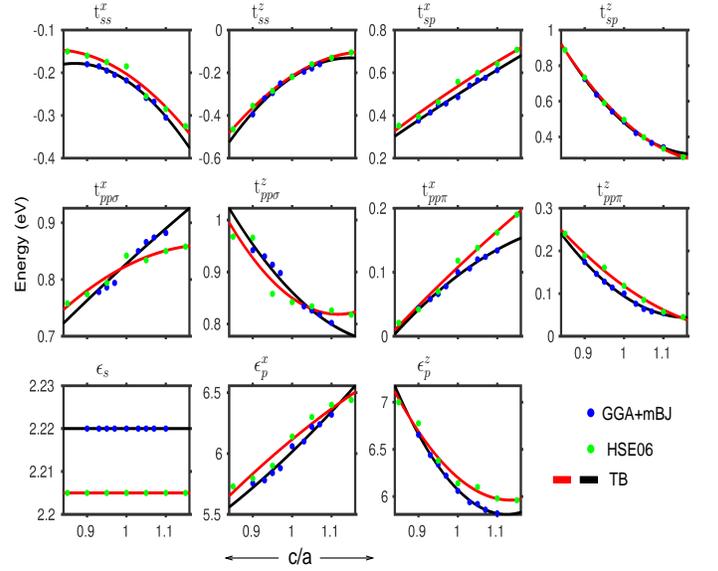}
\caption{Variation of TB parameters as function of c/a for GGA+mBJ and HSE06 XC functionals.  }
\label{fig7s}
\end{figure}

\begin{figure}
\centering
\includegraphics[angle=-0.0,origin=c,height=5.5cm,width=5.7cm]{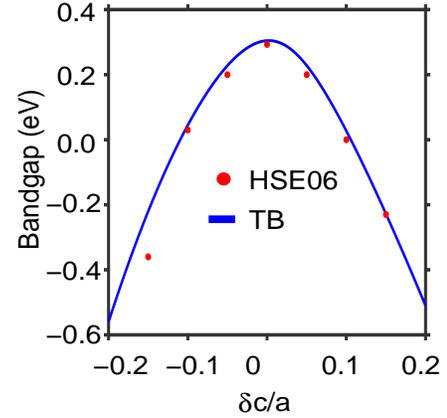}
\caption{ Same as Fig. \ref{fig8}a. However, the TB parameters are obtained by comparing the TB bands with the HSE06 based DFT bands.}
\label{fig8s}
\end{figure}

One of the widely known disadvantages of DFT calculation is the error associated with the XC functional as the latter affects the accuracy of the bandgap. Since the bandgap is a crucial parameter in examining the NI-TI phase transition, the validity of the latter needs to be checked for different types of XC functionals. In this appendix we have examined the validity, both quantitatively and qualitatively, of the phase transition  within the framework of four XC functionals, namely, GGA+mBJ(TB), GGA+mBJ(JTB)(This is the recent one incorporated in WEIN2k and is proposed to be appropriate for halide perovskites\cite{new_mBJ,new_mbj1}.), HSE06 (default mixing = 0.25), and HSE (mixing = 0.3). Here we demonstrate the case of $\alpha$-CsSnI$_3$ alone. Variation of the  bandgap as a function of c/a for these functionals is shown in Fig. \ref{fig6s}. We observe that while the bandgap varies reasonably with different XC functionals for the unstrained compound, with strain the variation weakens.  The critical strain (c/a) values to make a transition from the NI to TI state are found to be in the range 1.1 to 1.13 ( 0.87 to 0.89) for compressive (tensile) strain condition which we believe is practically feasible.

Irrespective of the minor quantitative deviations, the qualitative features of the band topology remain intact and not affected by the XC functionals. In Fig. \ref{fig7s}, we have shown the variation of the TB parameters for two XC functionals, GGA+mBJ and HSE06. Other than some constant shift in the onsite energy parameters ($\epsilon$ ), the variations in the interaction parameters are nearly identical. Taking the HSE06 optimized TB parameters, we have recalculated the bandgap as a function c/a and compared them with the HSE06 based DFT results in Fig. \ref{fig8s}. The excellent agreement shows that the proposed TB model is least affected by the type of XC functional.

\newpage

\bibliography{paper}

\begin{thebibliography}{49}%
\makeatletter
\providecommand \@ifxundefined [1]{%
 \@ifx{#1\undefined}
}%
\providecommand \@ifnum [1]{%
 \ifnum #1\expandafter \@firstoftwo
 \else \expandafter \@secondoftwo
 \fi
}%
\providecommand \@ifx [1]{%
 \ifx #1\expandafter \@firstoftwo
 \else \expandafter \@secondoftwo
 \fi
}%
\providecommand \natexlab [1]{#1}%
\providecommand \enquote  [1]{``#1''}%
\providecommand \bibnamefont  [1]{#1}%
\providecommand \bibfnamefont [1]{#1}%
\providecommand \citenamefont [1]{#1}%
\providecommand \href@noop [0]{\@secondoftwo}%
\providecommand \href [0]{\begingroup \@sanitize@url \@href}%
\providecommand \@href[1]{\@@startlink{#1}\@@href}%
\providecommand \@@href[1]{\endgroup#1\@@endlink}%
\providecommand \@sanitize@url [0]{\catcode `\\12\catcode `\$12\catcode
  `\&12\catcode `\#12\catcode `\^12\catcode `\_12\catcode `\%12\relax}%
\providecommand \@@startlink[1]{}%
\providecommand \@@endlink[0]{}%
\providecommand \url  [0]{\begingroup\@sanitize@url \@url }%
\providecommand \@url [1]{\endgroup\@href {#1}{\urlprefix }}%
\providecommand \urlprefix  [0]{URL }%
\providecommand \Eprint [0]{\href }%
\providecommand \doibase [0]{http://dx.doi.org/}%
\providecommand \selectlanguage [0]{\@gobble}%
\providecommand \bibinfo  [0]{\@secondoftwo}%
\providecommand \bibfield  [0]{\@secondoftwo}%
\providecommand \translation [1]{[#1]}%
\providecommand \BibitemOpen [0]{}%
\providecommand \bibitemStop [0]{}%
\providecommand \bibitemNoStop [0]{.\EOS\space}%
\providecommand \EOS [0]{\spacefactor3000\relax}%
\providecommand \BibitemShut  [1]{\csname bibitem#1\endcsname}%
\let\auto@bib@innerbib\@empty
%</preamble>
\bibitem [{\citenamefont {Kojima}\ \emph {et~al.}(2009)\citenamefont {Kojima},
  \citenamefont {Teshima}, \citenamefont {Shirai},\ and\ \citenamefont
  {Miyasaka}}]{pv0}%
  \BibitemOpen
  \bibfield  {author} {\bibinfo {author} {\bibfnamefont {A.}~\bibnamefont
  {Kojima}}, \bibinfo {author} {\bibfnamefont {K.}~\bibnamefont {Teshima}},
  \bibinfo {author} {\bibfnamefont {Y.}~\bibnamefont {Shirai}}, \ and\ \bibinfo
  {author} {\bibfnamefont {T.}~\bibnamefont {Miyasaka}},\ }\href {\doibase
  10.1021/ja809598r} {\bibfield  {journal} {\bibinfo  {journal} {Journal of the
  American Chemical Society}\ }\textbf {\bibinfo {volume} {131}},\ \bibinfo
  {pages} {6050} (\bibinfo {year} {2009})}\BibitemShut {NoStop}%
\bibitem [{\citenamefont {Berger}(2018)}]{pv01}%
  \BibitemOpen
  \bibfield  {author} {\bibinfo {author} {\bibfnamefont {R.~F.}\ \bibnamefont
  {Berger}},\ }\href {\doibase 10.1002/chem.201706126} {\bibfield  {journal}
  {\bibinfo  {journal} {Chemistry â€“ A European Journal}\ }\textbf {\bibinfo
  {volume} {24}},\ \bibinfo {pages} {8708} (\bibinfo {year}
  {2018})}\BibitemShut {NoStop}%
\bibitem [{\citenamefont {Qiu}\ \emph {et~al.}(2017)\citenamefont {Qiu},
  \citenamefont {Cao}, \citenamefont {Yuan}, \citenamefont {Chen},
  \citenamefont {Qiu}, \citenamefont {Jiang}, \citenamefont {Ye}, \citenamefont
  {Wang}, \citenamefont {Zeng}, \citenamefont {Liu},\ and\ \citenamefont
  {Kanatzidis}}]{pv1}%
  \BibitemOpen
  \bibfield  {author} {\bibinfo {author} {\bibfnamefont {X.}~\bibnamefont
  {Qiu}}, \bibinfo {author} {\bibfnamefont {B.}~\bibnamefont {Cao}}, \bibinfo
  {author} {\bibfnamefont {S.}~\bibnamefont {Yuan}}, \bibinfo {author}
  {\bibfnamefont {X.}~\bibnamefont {Chen}}, \bibinfo {author} {\bibfnamefont
  {Z.}~\bibnamefont {Qiu}}, \bibinfo {author} {\bibfnamefont {Y.}~\bibnamefont
  {Jiang}}, \bibinfo {author} {\bibfnamefont {Q.}~\bibnamefont {Ye}}, \bibinfo
  {author} {\bibfnamefont {H.}~\bibnamefont {Wang}}, \bibinfo {author}
  {\bibfnamefont {H.}~\bibnamefont {Zeng}}, \bibinfo {author} {\bibfnamefont
  {J.}~\bibnamefont {Liu}}, \ and\ \bibinfo {author} {\bibfnamefont {M.~G.}\
  \bibnamefont {Kanatzidis}},\ }\href {\doibase
  https://doi.org/10.1016/j.solmat.2016.09.022} {\bibfield  {journal} {\bibinfo
   {journal} {Solar Energy Materials and Solar Cells}\ }\textbf {\bibinfo
  {volume} {159}},\ \bibinfo {pages} {227 } (\bibinfo {year}
  {2017})}\BibitemShut {NoStop}%
\bibitem [{\citenamefont {Jiang}\ \emph {et~al.}(2018)\citenamefont {Jiang},
  \citenamefont {Onwudinanti}, \citenamefont {Hatton}, \citenamefont
  {Bobbert},\ and\ \citenamefont {Tao}}]{pv2}%
  \BibitemOpen
  \bibfield  {author} {\bibinfo {author} {\bibfnamefont {J.}~\bibnamefont
  {Jiang}}, \bibinfo {author} {\bibfnamefont {C.~K.}\ \bibnamefont
  {Onwudinanti}}, \bibinfo {author} {\bibfnamefont {R.~A.}\ \bibnamefont
  {Hatton}}, \bibinfo {author} {\bibfnamefont {P.~A.}\ \bibnamefont {Bobbert}},
  \ and\ \bibinfo {author} {\bibfnamefont {S.}~\bibnamefont {Tao}},\ }\href
  {\doibase 10.1021/acs.jpcc.8b04013} {\bibfield  {journal} {\bibinfo
  {journal} {The Journal of Physical Chemistry C}\ }\textbf {\bibinfo {volume}
  {122}},\ \bibinfo {pages} {17660} (\bibinfo {year} {2018})}\BibitemShut
  {NoStop}%
\bibitem [{\citenamefont {Yang}\ \emph {et~al.}(2012)\citenamefont {Yang},
  \citenamefont {Setyawan}, \citenamefont {Wang}, \citenamefont {{Buongiorno
  Nardelli}},\ and\ \citenamefont {Curtarolo}}]{Nature_mat}%
  \BibitemOpen
  \bibfield  {author} {\bibinfo {author} {\bibfnamefont {K.}~\bibnamefont
  {Yang}}, \bibinfo {author} {\bibfnamefont {W.}~\bibnamefont {Setyawan}},
  \bibinfo {author} {\bibfnamefont {S.}~\bibnamefont {Wang}}, \bibinfo {author}
  {\bibfnamefont {M.}~\bibnamefont {{Buongiorno Nardelli}}}, \ and\ \bibinfo
  {author} {\bibfnamefont {S.}~\bibnamefont {Curtarolo}},\ }\href {\doibase
  10.1038/nmat3332} {\bibfield  {journal} {\bibinfo  {journal} {Nature
  Materials}\ }\textbf {\bibinfo {volume} {11}},\ \bibinfo {pages} {614}
  (\bibinfo {year} {2012})}\BibitemShut {NoStop}%
\bibitem [{\citenamefont {Jung}\ \emph {et~al.}(2017)\citenamefont {Jung},
  \citenamefont {Lee}, \citenamefont {Walsh},\ and\ \citenamefont
  {Soon}}]{doping1}%
  \BibitemOpen
  \bibfield  {author} {\bibinfo {author} {\bibfnamefont {Y.-K.}\ \bibnamefont
  {Jung}}, \bibinfo {author} {\bibfnamefont {J.-H.}\ \bibnamefont {Lee}},
  \bibinfo {author} {\bibfnamefont {A.}~\bibnamefont {Walsh}}, \ and\ \bibinfo
  {author} {\bibfnamefont {A.}~\bibnamefont {Soon}},\ }\href {\doibase
  10.1021/acs.chemmater.7b00260} {\bibfield  {journal} {\bibinfo  {journal}
  {Chemistry of Materials}\ }\textbf {\bibinfo {volume} {29}},\ \bibinfo
  {pages} {3181} (\bibinfo {year} {2017})}\BibitemShut {NoStop}%
\bibitem [{\citenamefont {Shi}\ \emph {et~al.}(2015)\citenamefont {Shi},
  \citenamefont {Liu}, \citenamefont {Xu}, \citenamefont {Xiong}, \citenamefont
  {Wu},\ and\ \citenamefont {Duan}}]{doping2}%
  \BibitemOpen
  \bibfield  {author} {\bibinfo {author} {\bibfnamefont {W.-J.}\ \bibnamefont
  {Shi}}, \bibinfo {author} {\bibfnamefont {J.}~\bibnamefont {Liu}}, \bibinfo
  {author} {\bibfnamefont {Y.}~\bibnamefont {Xu}}, \bibinfo {author}
  {\bibfnamefont {S.-J.}\ \bibnamefont {Xiong}}, \bibinfo {author}
  {\bibfnamefont {J.}~\bibnamefont {Wu}}, \ and\ \bibinfo {author}
  {\bibfnamefont {W.}~\bibnamefont {Duan}},\ }\href {\doibase
  10.1103/PhysRevB.92.205118} {\bibfield  {journal} {\bibinfo  {journal} {Phys.
  Rev. B}\ }\textbf {\bibinfo {volume} {92}},\ \bibinfo {pages} {205118}
  (\bibinfo {year} {2015})}\BibitemShut {NoStop}%
\bibitem [{\citenamefont {Grote}\ and\ \citenamefont {Berger}(2015)}]{strain1}%
  \BibitemOpen
  \bibfield  {author} {\bibinfo {author} {\bibfnamefont {C.}~\bibnamefont
  {Grote}}\ and\ \bibinfo {author} {\bibfnamefont {R.~F.}\ \bibnamefont
  {Berger}},\ }\href {\doibase 10.1021/acs.jpcc.5b07446} {\bibfield  {journal}
  {\bibinfo  {journal} {The Journal of Physical Chemistry C}\ }\textbf
  {\bibinfo {volume} {119}},\ \bibinfo {pages} {22832} (\bibinfo {year}
  {2015})}\BibitemShut {NoStop}%
\bibitem [{\citenamefont {Jin}\ \emph {et~al.}(2012)\citenamefont {Jin},
  \citenamefont {Im},\ and\ \citenamefont {Freeman}}]{strain3}%
  \BibitemOpen
  \bibfield  {author} {\bibinfo {author} {\bibfnamefont {H.}~\bibnamefont
  {Jin}}, \bibinfo {author} {\bibfnamefont {J.}~\bibnamefont {Im}}, \ and\
  \bibinfo {author} {\bibfnamefont {A.~J.}\ \bibnamefont {Freeman}},\ }\href
  {\doibase 10.1103/PhysRevB.86.121102} {\bibfield  {journal} {\bibinfo
  {journal} {Phys. Rev. B}\ }\textbf {\bibinfo {volume} {86}},\ \bibinfo
  {pages} {121102} (\bibinfo {year} {2012})}\BibitemShut {NoStop}%
\bibitem [{\citenamefont {Song}\ \emph {et~al.}(2017)\citenamefont {Song},
  \citenamefont {Gao}, \citenamefont {Li},\ and\ \citenamefont
  {Zhang}}]{strain4}%
  \BibitemOpen
  \bibfield  {author} {\bibinfo {author} {\bibfnamefont {G.}~\bibnamefont
  {Song}}, \bibinfo {author} {\bibfnamefont {B.}~\bibnamefont {Gao}}, \bibinfo
  {author} {\bibfnamefont {G.}~\bibnamefont {Li}}, \ and\ \bibinfo {author}
  {\bibfnamefont {J.}~\bibnamefont {Zhang}},\ }\href {\doibase
  10.1039/C7RA07735A} {\bibfield  {journal} {\bibinfo  {journal} {RSC Adv.}\
  }\textbf {\bibinfo {volume} {7}},\ \bibinfo {pages} {41077} (\bibinfo {year}
  {2017})}\BibitemShut {NoStop}%
\bibitem [{\citenamefont {Yu}\ \emph {et~al.}(2011)\citenamefont {Yu},
  \citenamefont {Chen}, \citenamefont {J.~Wang}, \citenamefont {Pfenninger},
  \citenamefont {Vockic}, \citenamefont {Kenney},\ and\ \citenamefont
  {Shum}}]{temp}%
  \BibitemOpen
  \bibfield  {author} {\bibinfo {author} {\bibfnamefont {C.}~\bibnamefont
  {Yu}}, \bibinfo {author} {\bibfnamefont {Z.}~\bibnamefont {Chen}}, \bibinfo
  {author} {\bibfnamefont {J.}~\bibnamefont {J.~Wang}}, \bibinfo {author}
  {\bibfnamefont {W.}~\bibnamefont {Pfenninger}}, \bibinfo {author}
  {\bibfnamefont {N.}~\bibnamefont {Vockic}}, \bibinfo {author} {\bibfnamefont
  {J.~T.}\ \bibnamefont {Kenney}}, \ and\ \bibinfo {author} {\bibfnamefont
  {K.}~\bibnamefont {Shum}},\ }\href {\doibase 10.1063/1.3638699} {\bibfield
  {journal} {\bibinfo  {journal} {Journal of Applied Physics}\ }\textbf
  {\bibinfo {volume} {110}},\ \bibinfo {pages} {063526} (\bibinfo {year}
  {2011})}\BibitemShut {NoStop}%
\bibitem [{\citenamefont {Gou}\ \emph {et~al.}(2017)\citenamefont {Gou},
  \citenamefont {Young}, \citenamefont {Liu},\ and\ \citenamefont
  {Rondinelli}}]{12}%
  \BibitemOpen
  \bibfield  {author} {\bibinfo {author} {\bibfnamefont {G.}~\bibnamefont
  {Gou}}, \bibinfo {author} {\bibfnamefont {J.}~\bibnamefont {Young}}, \bibinfo
  {author} {\bibfnamefont {X.}~\bibnamefont {Liu}}, \ and\ \bibinfo {author}
  {\bibfnamefont {J.~M.}\ \bibnamefont {Rondinelli}},\ }\href {\doibase
  10.1021/acs.inorgchem.6b01701} {\bibfield  {journal} {\bibinfo  {journal}
  {Inorganic Chemistry}\ }\textbf {\bibinfo {volume} {56}},\ \bibinfo {pages}
  {26} (\bibinfo {year} {2017})},\ \bibinfo {note} {pMID: 27682844}\BibitemShut
  {NoStop}%
\bibitem [{\citenamefont {Kashikar}\ \emph {et~al.}(2018)\citenamefont
  {Kashikar}, \citenamefont {Khamari},\ and\ \citenamefont {Nanda}}]{Ravi}%
  \BibitemOpen
  \bibfield  {author} {\bibinfo {author} {\bibfnamefont {R.}~\bibnamefont
  {Kashikar}}, \bibinfo {author} {\bibfnamefont {B.}~\bibnamefont {Khamari}}, \
  and\ \bibinfo {author} {\bibfnamefont {B.~R.~K.}\ \bibnamefont {Nanda}},\
  }\href {\doibase 10.1103/PhysRevMaterials.2.124204} {\bibfield  {journal}
  {\bibinfo  {journal} {Phys. Rev. Materials}\ }\textbf {\bibinfo {volume}
  {2}},\ \bibinfo {pages} {124204} (\bibinfo {year} {2018})}\BibitemShut
  {NoStop}%
\bibitem [{\citenamefont {Frohna}\ \emph {et~al.}(2018)\citenamefont {Frohna},
  \citenamefont {Deshpande}, \citenamefont {Harter}, \citenamefont {Peng},
  \citenamefont {Barker}, \citenamefont {Neaton}, \citenamefont {Louie},
  \citenamefont {Bakr}, \citenamefont {Hsieh},\ and\ \citenamefont
  {Bernardi}}]{rashba_soc}%
  \BibitemOpen
  \bibfield  {author} {\bibinfo {author} {\bibfnamefont {K.}~\bibnamefont
  {Frohna}}, \bibinfo {author} {\bibfnamefont {T.}~\bibnamefont {Deshpande}},
  \bibinfo {author} {\bibfnamefont {J.}~\bibnamefont {Harter}}, \bibinfo
  {author} {\bibfnamefont {W.}~\bibnamefont {Peng}}, \bibinfo {author}
  {\bibfnamefont {B.~A.}\ \bibnamefont {Barker}}, \bibinfo {author}
  {\bibfnamefont {J.~B.}\ \bibnamefont {Neaton}}, \bibinfo {author}
  {\bibfnamefont {S.~G.}\ \bibnamefont {Louie}}, \bibinfo {author}
  {\bibfnamefont {O.~M.}\ \bibnamefont {Bakr}}, \bibinfo {author}
  {\bibfnamefont {D.}~\bibnamefont {Hsieh}}, \ and\ \bibinfo {author}
  {\bibfnamefont {M.}~\bibnamefont {Bernardi}},\ }\href
  {http://dx.doi.org/10.1038/s41467-018-04212-w} {\bibfield  {journal}
  {\bibinfo  {journal} {Nature Communications}\ }\textbf {\bibinfo {volume}
  {9}} (\bibinfo {year} {2018})}\BibitemShut {NoStop}%
\bibitem [{\citenamefont {Kim}\ \emph {et~al.}(2014)\citenamefont {Kim},
  \citenamefont {Im}, \citenamefont {Freeman}, \citenamefont {Ihm},\ and\
  \citenamefont {Jin}}]{rashba_soc_1}%
  \BibitemOpen
  \bibfield  {author} {\bibinfo {author} {\bibfnamefont {M.}~\bibnamefont
  {Kim}}, \bibinfo {author} {\bibfnamefont {J.}~\bibnamefont {Im}}, \bibinfo
  {author} {\bibfnamefont {A.~J.}\ \bibnamefont {Freeman}}, \bibinfo {author}
  {\bibfnamefont {J.}~\bibnamefont {Ihm}}, \ and\ \bibinfo {author}
  {\bibfnamefont {H.}~\bibnamefont {Jin}},\ }\href {\doibase
  10.1073/pnas.1405780111} {\bibfield  {journal} {\bibinfo  {journal}
  {Proceedings of the National Academy of Sciences}\ }\textbf {\bibinfo
  {volume} {111}},\ \bibinfo {pages} {6900} (\bibinfo {year}
  {2014})}\BibitemShut {NoStop}%
\bibitem [{\citenamefont {Tao}\ \emph {et~al.}(2017)\citenamefont {Tao},
  \citenamefont {Cao},\ and\ \citenamefont {Bobbert}}]{DFT_1_2}%
  \BibitemOpen
  \bibfield  {author} {\bibinfo {author} {\bibfnamefont {S.~X.}\ \bibnamefont
  {Tao}}, \bibinfo {author} {\bibfnamefont {X.}~\bibnamefont {Cao}}, \ and\
  \bibinfo {author} {\bibfnamefont {P.~A.}\ \bibnamefont {Bobbert}},\ }\href
  {\doibase 10.1038/s41598-017-14435-4} {\bibfield  {journal} {\bibinfo
  {journal} {Scientific Reports}\ }\textbf {\bibinfo {volume} {7}},\ \bibinfo
  {pages} {1} (\bibinfo {year} {2017})}\BibitemShut {NoStop}%
\bibitem [{\citenamefont {Yu}\ \emph {et~al.}(2013)\citenamefont {Yu},
  \citenamefont {Ren}, \citenamefont {Chen},\ and\ \citenamefont
  {Shum}}]{struct1}%
  \BibitemOpen
  \bibfield  {author} {\bibinfo {author} {\bibfnamefont {C.}~\bibnamefont
  {Yu}}, \bibinfo {author} {\bibfnamefont {Y.}~\bibnamefont {Ren}}, \bibinfo
  {author} {\bibfnamefont {Z.}~\bibnamefont {Chen}}, \ and\ \bibinfo {author}
  {\bibfnamefont {K.}~\bibnamefont {Shum}},\ }\href {\doibase
  10.1063/1.4826068} {\bibfield  {journal} {\bibinfo  {journal} {Journal of
  Applied Physics}\ }\textbf {\bibinfo {volume} {114}},\ \bibinfo {pages}
  {163505} (\bibinfo {year} {2013})}\BibitemShut {NoStop}%
\bibitem [{\citenamefont {Yang}\ \emph {et~al.}(2017)\citenamefont {Yang},
  \citenamefont {Skelton}, \citenamefont {da~Silva}, \citenamefont {Frost},\
  and\ \citenamefont {Walsh}}]{struct2}%
  \BibitemOpen
  \bibfield  {author} {\bibinfo {author} {\bibfnamefont {R.~X.}\ \bibnamefont
  {Yang}}, \bibinfo {author} {\bibfnamefont {J.~M.}\ \bibnamefont {Skelton}},
  \bibinfo {author} {\bibfnamefont {E.~L.}\ \bibnamefont {da~Silva}}, \bibinfo
  {author} {\bibfnamefont {J.~M.}\ \bibnamefont {Frost}}, \ and\ \bibinfo
  {author} {\bibfnamefont {A.}~\bibnamefont {Walsh}},\ }\href {\doibase
  10.1021/acs.jpclett.7b02423} {\bibfield  {journal} {\bibinfo  {journal} {The
  Journal of Physical Chemistry Letters}\ }\textbf {\bibinfo {volume} {8}},\
  \bibinfo {pages} {4720} (\bibinfo {year} {2017})}\BibitemShut {NoStop}%
\bibitem [{\citenamefont {Liu}\ \emph {et~al.}(2016)\citenamefont {Liu},
  \citenamefont {Kim}, \citenamefont {Tan},\ and\ \citenamefont
  {Rappe}}]{Nano_letter}%
  \BibitemOpen
  \bibfield  {author} {\bibinfo {author} {\bibfnamefont {S.}~\bibnamefont
  {Liu}}, \bibinfo {author} {\bibfnamefont {Y.}~\bibnamefont {Kim}}, \bibinfo
  {author} {\bibfnamefont {L.~Z.}\ \bibnamefont {Tan}}, \ and\ \bibinfo
  {author} {\bibfnamefont {A.~M.}\ \bibnamefont {Rappe}},\ }\href@noop {}
  {\bibfield  {journal} {\bibinfo  {journal} {Nano Letters}\ }\textbf {\bibinfo
  {volume} {16}},\ \bibinfo {pages} {1663} (\bibinfo {year}
  {2016})}\BibitemShut {NoStop}%
\bibitem [{\citenamefont {Boyer-Richard}\ \emph {et~al.}(2016)\citenamefont
  {Boyer-Richard}, \citenamefont {Katan}, \citenamefont {TraorÃ©},
  \citenamefont {Scholz}, \citenamefont {Jancu},\ and\ \citenamefont
  {Even}}]{JPCL}%
  \BibitemOpen
  \bibfield  {author} {\bibinfo {author} {\bibfnamefont {S.}~\bibnamefont
  {Boyer-Richard}}, \bibinfo {author} {\bibfnamefont {C.}~\bibnamefont
  {Katan}}, \bibinfo {author} {\bibfnamefont {B.}~\bibnamefont {TraorÃ©}},
  \bibinfo {author} {\bibfnamefont {R.}~\bibnamefont {Scholz}}, \bibinfo
  {author} {\bibfnamefont {J.-M.}\ \bibnamefont {Jancu}}, \ and\ \bibinfo
  {author} {\bibfnamefont {J.}~\bibnamefont {Even}},\ }\href@noop {} {\bibfield
   {journal} {\bibinfo  {journal} {The Journal of Physical Chemistry Letters}\
  }\textbf {\bibinfo {volume} {7}},\ \bibinfo {pages} {3833} (\bibinfo {year}
  {2016})}\BibitemShut {NoStop}%
\bibitem [{\citenamefont {Khamari}\ \emph {et~al.}(2018)\citenamefont
  {Khamari}, \citenamefont {Kashikar},\ and\ \citenamefont {Nanda}}]{BK}%
  \BibitemOpen
  \bibfield  {author} {\bibinfo {author} {\bibfnamefont {B.}~\bibnamefont
  {Khamari}}, \bibinfo {author} {\bibfnamefont {R.}~\bibnamefont {Kashikar}}, \
  and\ \bibinfo {author} {\bibfnamefont {B.~R.~K.}\ \bibnamefont {Nanda}},\
  }\href {\doibase 10.1103/PhysRevB.97.045149} {\bibfield  {journal} {\bibinfo
  {journal} {Phys. Rev. B}\ }\textbf {\bibinfo {volume} {97}},\ \bibinfo
  {pages} {045149} (\bibinfo {year} {2018})}\BibitemShut {NoStop}%
\bibitem [{\citenamefont {Chung}\ \emph {et~al.}(2012)\citenamefont {Chung},
  \citenamefont {Song}, \citenamefont {Im}, \citenamefont {Androulakis},
  \citenamefont {Malliakas}, \citenamefont {Li}, \citenamefont {Freeman},
  \citenamefont {Kenney},\ and\ \citenamefont {Kanatzidis}}]{struct3}%
  \BibitemOpen
  \bibfield  {author} {\bibinfo {author} {\bibfnamefont {I.}~\bibnamefont
  {Chung}}, \bibinfo {author} {\bibfnamefont {J.-H.}\ \bibnamefont {Song}},
  \bibinfo {author} {\bibfnamefont {J.}~\bibnamefont {Im}}, \bibinfo {author}
  {\bibfnamefont {J.}~\bibnamefont {Androulakis}}, \bibinfo {author}
  {\bibfnamefont {C.~D.}\ \bibnamefont {Malliakas}}, \bibinfo {author}
  {\bibfnamefont {H.}~\bibnamefont {Li}}, \bibinfo {author} {\bibfnamefont
  {A.~J.}\ \bibnamefont {Freeman}}, \bibinfo {author} {\bibfnamefont {J.~T.}\
  \bibnamefont {Kenney}}, \ and\ \bibinfo {author} {\bibfnamefont {M.~G.}\
  \bibnamefont {Kanatzidis}},\ }\href {\doibase 10.1021/ja301539s} {\bibfield
  {journal} {\bibinfo  {journal} {Journal of the American Chemical Society}\
  }\textbf {\bibinfo {volume} {134}},\ \bibinfo {pages} {8579} (\bibinfo {year}
  {2012})}\BibitemShut {NoStop}%
\bibitem [{\citenamefont {Hamann}(1979)}]{LAPW}%
  \BibitemOpen
  \bibfield  {author} {\bibinfo {author} {\bibfnamefont {D.~R.}\ \bibnamefont
  {Hamann}},\ }\href {\doibase 10.1103/PhysRevLett.42.662} {\bibfield
  {journal} {\bibinfo  {journal} {Phys. Rev. Lett.}\ }\textbf {\bibinfo
  {volume} {42}},\ \bibinfo {pages} {662} (\bibinfo {year} {1979})}\BibitemShut
  {NoStop}%
\bibitem [{\citenamefont {Blaha}\ \emph {et~al.}(2001)\citenamefont {Blaha},
  \citenamefont {Schwartz}, \citenamefont {Madsen}, \citenamefont {Kvasnicka},\
  and\ \citenamefont {Luitz}}]{Blaha}%
  \BibitemOpen
  \bibfield  {author} {\bibinfo {author} {\bibfnamefont {P.}~\bibnamefont
  {Blaha}}, \bibinfo {author} {\bibfnamefont {K.}~\bibnamefont {Schwartz}},
  \bibinfo {author} {\bibfnamefont {G.}~\bibnamefont {Madsen}}, \bibinfo
  {author} {\bibfnamefont {D.}~\bibnamefont {Kvasnicka}}, \ and\ \bibinfo
  {author} {\bibfnamefont {J.}~\bibnamefont {Luitz}},\ }\href@noop {} {\emph
  {\bibinfo {title} {WIEN2k An Augmanted Plane Wave+Local Orbitals Program for
  Calculating Crystal Properties}}}\ (\bibinfo  {publisher} {Karlheinz
  Schwartz, Tech. Universitt Wien, Austria},\ \bibinfo {year}
  {2001})\BibitemShut {NoStop}%
\bibitem [{\citenamefont {Perdew}\ \emph {et~al.}(1996)\citenamefont {Perdew},
  \citenamefont {Burke},\ and\ \citenamefont {Ernzerhof}}]{GGA}%
  \BibitemOpen
  \bibfield  {author} {\bibinfo {author} {\bibfnamefont {J.~P.}\ \bibnamefont
  {Perdew}}, \bibinfo {author} {\bibfnamefont {K.}~\bibnamefont {Burke}}, \
  and\ \bibinfo {author} {\bibfnamefont {M.}~\bibnamefont {Ernzerhof}},\ }\href
  {\doibase 10.1103/PhysRevLett.77.3865} {\bibfield  {journal} {\bibinfo
  {journal} {Phys. Rev. Lett.}\ }\textbf {\bibinfo {volume} {77}},\ \bibinfo
  {pages} {3865} (\bibinfo {year} {1996})}\BibitemShut {NoStop}%
\bibitem [{\citenamefont {Tran}\ and\ \citenamefont {Blaha}(2009)}]{mbj-1}%
  \BibitemOpen
  \bibfield  {author} {\bibinfo {author} {\bibfnamefont {F.}~\bibnamefont
  {Tran}}\ and\ \bibinfo {author} {\bibfnamefont {P.}~\bibnamefont {Blaha}},\
  }\href {\doibase 10.1103/PhysRevLett.102.226401} {\bibfield  {journal}
  {\bibinfo  {journal} {Phys. Rev. Lett.}\ }\textbf {\bibinfo {volume} {102}},\
  \bibinfo {pages} {226401} (\bibinfo {year} {2009})}\BibitemShut {NoStop}%
\bibitem [{\citenamefont {Tran}\ \emph {et~al.}(2007)\citenamefont {Tran},
  \citenamefont {Blaha},\ and\ \citenamefont {Schwarz}}]{mbj-2}%
  \BibitemOpen
  \bibfield  {author} {\bibinfo {author} {\bibfnamefont {F.}~\bibnamefont
  {Tran}}, \bibinfo {author} {\bibfnamefont {P.}~\bibnamefont {Blaha}}, \ and\
  \bibinfo {author} {\bibfnamefont {K.}~\bibnamefont {Schwarz}},\ }\href
  {http://stacks.iop.org/0953-8984/19/i=19/a=196208} {\bibfield  {journal}
  {\bibinfo  {journal} {Journal of Physics: Condensed Matter}\ }\textbf
  {\bibinfo {volume} {19}},\ \bibinfo {pages} {196208} (\bibinfo {year}
  {2007})}\BibitemShut {NoStop}%
\bibitem [{\citenamefont {Borriello}\ \emph {et~al.}(2008)\citenamefont
  {Borriello}, \citenamefont {Cantele},\ and\ \citenamefont
  {Ninno}}]{PRB_struct}%
  \BibitemOpen
  \bibfield  {author} {\bibinfo {author} {\bibfnamefont {I.}~\bibnamefont
  {Borriello}}, \bibinfo {author} {\bibfnamefont {G.}~\bibnamefont {Cantele}},
  \ and\ \bibinfo {author} {\bibfnamefont {D.}~\bibnamefont {Ninno}},\ }\href
  {\doibase 10.1103/PhysRevB.77.235214} {\bibfield  {journal} {\bibinfo
  {journal} {Phys. Rev. B}\ }\textbf {\bibinfo {volume} {77}},\ \bibinfo
  {pages} {235214} (\bibinfo {year} {2008})}\BibitemShut {NoStop}%
\bibitem [{\citenamefont {Huang}\ and\ \citenamefont
  {Lambrecht}(2013)}]{Lamb_1}%
  \BibitemOpen
  \bibfield  {author} {\bibinfo {author} {\bibfnamefont {L.-y.}\ \bibnamefont
  {Huang}}\ and\ \bibinfo {author} {\bibfnamefont {W.~R.~L.}\ \bibnamefont
  {Lambrecht}},\ }\href {\doibase 10.1103/PhysRevB.88.165203} {\bibfield
  {journal} {\bibinfo  {journal} {Phys. Rev. B}\ }\textbf {\bibinfo {volume}
  {88}},\ \bibinfo {pages} {165203} (\bibinfo {year} {2013})}\BibitemShut
  {NoStop}%
\bibitem [{\citenamefont {Huang}\ and\ \citenamefont
  {Lambrecht}(2016)}]{Lamb_2}%
  \BibitemOpen
  \bibfield  {author} {\bibinfo {author} {\bibfnamefont {L.-y.}\ \bibnamefont
  {Huang}}\ and\ \bibinfo {author} {\bibfnamefont {W.~R.~L.}\ \bibnamefont
  {Lambrecht}},\ }\href {\doibase 10.1103/PhysRevB.93.195211} {\bibfield
  {journal} {\bibinfo  {journal} {Phys. Rev. B}\ }\textbf {\bibinfo {volume}
  {93}},\ \bibinfo {pages} {195211} (\bibinfo {year} {2016})}\BibitemShut
  {NoStop}%
\bibitem [{\citenamefont {Even}\ \emph {et~al.}(2014)\citenamefont {Even},
  \citenamefont {Pedesseau}, \citenamefont {Jancu},\ and\ \citenamefont
  {Katan}}]{band_1}%
  \BibitemOpen
  \bibfield  {author} {\bibinfo {author} {\bibfnamefont {J.}~\bibnamefont
  {Even}}, \bibinfo {author} {\bibfnamefont {L.}~\bibnamefont {Pedesseau}},
  \bibinfo {author} {\bibfnamefont {J.-M.}\ \bibnamefont {Jancu}}, \ and\
  \bibinfo {author} {\bibfnamefont {C.}~\bibnamefont {Katan}},\ }\href@noop {}
  {\bibfield  {journal} {\bibinfo  {journal} {physica status solidi (RRL) â€“
  Rapid Research Letters}\ }\textbf {\bibinfo {volume} {8}} (\bibinfo {year}
  {2014})}\BibitemShut {NoStop}%
\bibitem [{\citenamefont {Tai}\ \emph {et~al.}(2019)\citenamefont {Tai},
  \citenamefont {Tang},\ and\ \citenamefont {Yan}}]{eg_eff}%
  \BibitemOpen
  \bibfield  {author} {\bibinfo {author} {\bibfnamefont {Q.}~\bibnamefont
  {Tai}}, \bibinfo {author} {\bibfnamefont {K.-C.}\ \bibnamefont {Tang}}, \
  and\ \bibinfo {author} {\bibfnamefont {F.}~\bibnamefont {Yan}},\ }\href
  {\doibase 10.1039/C9EE01479A} {\bibfield  {journal} {\bibinfo  {journal}
  {Energy Environ. Sci.}\ }\textbf {\bibinfo {volume} {12}},\ \bibinfo {pages}
  {2375} (\bibinfo {year} {2019})}\BibitemShut {NoStop}%
\bibitem [{\citenamefont {Slater}\ and\ \citenamefont {Koster}(1954)}]{slater}%
  \BibitemOpen
  \bibfield  {author} {\bibinfo {author} {\bibfnamefont {J.~C.}\ \bibnamefont
  {Slater}}\ and\ \bibinfo {author} {\bibfnamefont {G.~F.}\ \bibnamefont
  {Koster}},\ }\href {\doibase 10.1103/PhysRev.94.1498} {\bibfield  {journal}
  {\bibinfo  {journal} {Phys. Rev.}\ }\textbf {\bibinfo {volume} {94}},\
  \bibinfo {pages} {1498} (\bibinfo {year} {1954})}\BibitemShut {NoStop}%
\bibitem [{\citenamefont {AsbÃ³th}\ \emph {et~al.}(2016)\citenamefont
  {AsbÃ³th}, \citenamefont {OroszlÃ¡ny},\ and\ \citenamefont {PÃ¡lyi}}]{Book}%
  \BibitemOpen
  \bibfield  {author} {\bibinfo {author} {\bibfnamefont {J.~K.}\ \bibnamefont
  {AsbÃ³th}}, \bibinfo {author} {\bibfnamefont {L.}~\bibnamefont {OroszlÃ¡ny}},
  \ and\ \bibinfo {author} {\bibfnamefont {A.}~\bibnamefont {PÃ¡lyi}},\ }\href
  {\doibase 10.1007/978-3-319-25607-8} {\bibfield  {journal} {\bibinfo
  {journal} {Lecture Notes in Physics}\ } (\bibinfo {year} {2016}),\
  10.1007/978-3-319-25607-8}\BibitemShut {NoStop}%
\bibitem [{\citenamefont {Lee}\ \emph {et~al.}(2010)\citenamefont {Lee},
  \citenamefont {Arovas},\ and\ \citenamefont {Wu}}]{sro}%
  \BibitemOpen
  \bibfield  {author} {\bibinfo {author} {\bibfnamefont {W.-C.}\ \bibnamefont
  {Lee}}, \bibinfo {author} {\bibfnamefont {D.~P.}\ \bibnamefont {Arovas}}, \
  and\ \bibinfo {author} {\bibfnamefont {C.}~\bibnamefont {Wu}},\ }\href
  {\doibase 10.1103/PhysRevB.81.184403} {\bibfield  {journal} {\bibinfo
  {journal} {Phys. Rev. B}\ }\textbf {\bibinfo {volume} {81}},\ \bibinfo
  {pages} {184403} (\bibinfo {year} {2010})}\BibitemShut {NoStop}%
\bibitem [{\citenamefont {Trots}\ and\ \citenamefont {Myagkota}(2008)}]{1}%
  \BibitemOpen
  \bibfield  {author} {\bibinfo {author} {\bibfnamefont {D.}~\bibnamefont
  {Trots}}\ and\ \bibinfo {author} {\bibfnamefont {S.}~\bibnamefont
  {Myagkota}},\ }\href {\doibase https://doi.org/10.1016/j.jpcs.2008.05.007}
  {\bibfield  {journal} {\bibinfo  {journal} {Journal of Physics and Chemistry
  of Solids}\ }\textbf {\bibinfo {volume} {69}},\ \bibinfo {pages} {2520 }
  (\bibinfo {year} {2008})}\BibitemShut {NoStop}%
\bibitem [{\citenamefont {Fujii}\ \emph {et~al.}(1974)\citenamefont {Fujii},
  \citenamefont {Hoshino}, \citenamefont {Yamada},\ and\ \citenamefont
  {Shirane}}]{2}%
  \BibitemOpen
  \bibfield  {author} {\bibinfo {author} {\bibfnamefont {Y.}~\bibnamefont
  {Fujii}}, \bibinfo {author} {\bibfnamefont {S.}~\bibnamefont {Hoshino}},
  \bibinfo {author} {\bibfnamefont {Y.}~\bibnamefont {Yamada}}, \ and\ \bibinfo
  {author} {\bibfnamefont {G.}~\bibnamefont {Shirane}},\ }\href {\doibase
  10.1103/PhysRevB.9.4549} {\bibfield  {journal} {\bibinfo  {journal} {Phys.
  Rev. B}\ }\textbf {\bibinfo {volume} {9}},\ \bibinfo {pages} {4549} (\bibinfo
  {year} {1974})}\BibitemShut {NoStop}%
\bibitem [{\citenamefont {Hirotsu}\ \emph {et~al.}(1974)\citenamefont
  {Hirotsu}, \citenamefont {Harada}, \citenamefont {Iizumi},\ and\
  \citenamefont {Gesi}}]{3}%
  \BibitemOpen
  \bibfield  {author} {\bibinfo {author} {\bibfnamefont {S.}~\bibnamefont
  {Hirotsu}}, \bibinfo {author} {\bibfnamefont {J.}~\bibnamefont {Harada}},
  \bibinfo {author} {\bibfnamefont {M.}~\bibnamefont {Iizumi}}, \ and\ \bibinfo
  {author} {\bibfnamefont {K.}~\bibnamefont {Gesi}},\ }\href {\doibase
  10.1143/JPSJ.37.1393} {\bibfield  {journal} {\bibinfo  {journal} {Journal of
  the Physical Society of Japan}\ }\textbf {\bibinfo {volume} {37}},\ \bibinfo
  {pages} {1393} (\bibinfo {year} {1974})}\BibitemShut {NoStop}%
\bibitem [{\citenamefont {Thiele}\ \emph {et~al.}(1987)\citenamefont {Thiele},
  \citenamefont {Rotter},\ and\ \citenamefont {Schmidt}}]{6}%
  \BibitemOpen
  \bibfield  {author} {\bibinfo {author} {\bibfnamefont {G.}~\bibnamefont
  {Thiele}}, \bibinfo {author} {\bibfnamefont {H.~W.}\ \bibnamefont {Rotter}},
  \ and\ \bibinfo {author} {\bibfnamefont {K.~D.}\ \bibnamefont {Schmidt}},\
  }\href {\doibase 10.1002/zaac.19875450217} {\bibfield  {journal} {\bibinfo
  {journal} {Zeitschrift fÃ¼r anorganische und allgemeine Chemie}\ }\textbf
  {\bibinfo {volume} {545}},\ \bibinfo {pages} {148} (\bibinfo {year}
  {1987})}\BibitemShut {NoStop}%
\bibitem [{\citenamefont {Poglitsch}\ and\ \citenamefont {Weber}(1987)}]{7}%
  \BibitemOpen
  \bibfield  {author} {\bibinfo {author} {\bibfnamefont {A.}~\bibnamefont
  {Poglitsch}}\ and\ \bibinfo {author} {\bibfnamefont {D.}~\bibnamefont
  {Weber}},\ }\href {\doibase 10.1063/1.453467} {\bibfield  {journal} {\bibinfo
   {journal} {The Journal of Chemical Physics}\ }\textbf {\bibinfo {volume}
  {87}},\ \bibinfo {pages} {6373} (\bibinfo {year} {1987})}\BibitemShut
  {NoStop}%
\bibitem [{\citenamefont {Zhao}\ \emph {et~al.}(2018)\citenamefont {Zhao},
  \citenamefont {Jin}, \citenamefont {Huang}, \citenamefont {Liu},
  \citenamefont {Ma}, \citenamefont {Xue}, \citenamefont {Han}, \citenamefont
  {Ding}, \citenamefont {Ge}, \citenamefont {Feng},\ and\ \citenamefont
  {Hu}}]{10}%
  \BibitemOpen
  \bibfield  {author} {\bibinfo {author} {\bibfnamefont {B.}~\bibnamefont
  {Zhao}}, \bibinfo {author} {\bibfnamefont {S.-F.}\ \bibnamefont {Jin}},
  \bibinfo {author} {\bibfnamefont {S.}~\bibnamefont {Huang}}, \bibinfo
  {author} {\bibfnamefont {N.}~\bibnamefont {Liu}}, \bibinfo {author}
  {\bibfnamefont {J.-Y.}\ \bibnamefont {Ma}}, \bibinfo {author} {\bibfnamefont
  {D.-J.}\ \bibnamefont {Xue}}, \bibinfo {author} {\bibfnamefont
  {Q.}~\bibnamefont {Han}}, \bibinfo {author} {\bibfnamefont {J.}~\bibnamefont
  {Ding}}, \bibinfo {author} {\bibfnamefont {Q.-Q.}\ \bibnamefont {Ge}},
  \bibinfo {author} {\bibfnamefont {Y.}~\bibnamefont {Feng}}, \ and\ \bibinfo
  {author} {\bibfnamefont {J.-S.}\ \bibnamefont {Hu}},\ }\href {\doibase
  10.1021/jacs.8b06050} {\bibfield  {journal} {\bibinfo  {journal} {Journal of
  the American Chemical Society}\ }\textbf {\bibinfo {volume} {140}},\ \bibinfo
  {pages} {11716} (\bibinfo {year} {2018})}\BibitemShut {NoStop}%
\bibitem [{\citenamefont {Stoumpos}\ and\ \citenamefont
  {Kanatzidis}(2015)}]{11}%
  \BibitemOpen
  \bibfield  {author} {\bibinfo {author} {\bibfnamefont {C.~C.}\ \bibnamefont
  {Stoumpos}}\ and\ \bibinfo {author} {\bibfnamefont {M.~G.}\ \bibnamefont
  {Kanatzidis}},\ }\href {\doibase 10.1021/acs.accounts.5b00229} {\bibfield
  {journal} {\bibinfo  {journal} {Accounts of Chemical Research}\ }\textbf
  {\bibinfo {volume} {48}},\ \bibinfo {pages} {2791} (\bibinfo {year}
  {2015})}\BibitemShut {NoStop}%
\bibitem [{\citenamefont {Fabini}\ \emph {et~al.}(2016)\citenamefont {Fabini},
  \citenamefont {Laurita}, \citenamefont {Bechtel}, \citenamefont {Stoumpos},
  \citenamefont {Evans}, \citenamefont {Kontos}, \citenamefont {Raptis},
  \citenamefont {Falaras}, \citenamefont {Van~der Ven}, \citenamefont
  {Kanatzidis},\ and\ \citenamefont {Seshadri}}]{13}%
  \BibitemOpen
  \bibfield  {author} {\bibinfo {author} {\bibfnamefont {D.~H.}\ \bibnamefont
  {Fabini}}, \bibinfo {author} {\bibfnamefont {G.}~\bibnamefont {Laurita}},
  \bibinfo {author} {\bibfnamefont {J.~S.}\ \bibnamefont {Bechtel}}, \bibinfo
  {author} {\bibfnamefont {C.~C.}\ \bibnamefont {Stoumpos}}, \bibinfo {author}
  {\bibfnamefont {H.~A.}\ \bibnamefont {Evans}}, \bibinfo {author}
  {\bibfnamefont {A.~G.}\ \bibnamefont {Kontos}}, \bibinfo {author}
  {\bibfnamefont {Y.~S.}\ \bibnamefont {Raptis}}, \bibinfo {author}
  {\bibfnamefont {P.}~\bibnamefont {Falaras}}, \bibinfo {author} {\bibfnamefont
  {A.}~\bibnamefont {Van~der Ven}}, \bibinfo {author} {\bibfnamefont {M.~G.}\
  \bibnamefont {Kanatzidis}}, \ and\ \bibinfo {author} {\bibfnamefont
  {R.}~\bibnamefont {Seshadri}},\ }\href {\doibase 10.1021/jacs.6b06287}
  {\bibfield  {journal} {\bibinfo  {journal} {Journal of the American Chemical
  Society}\ }\textbf {\bibinfo {volume} {138}},\ \bibinfo {pages} {11820}
  (\bibinfo {year} {2016})}\BibitemShut {NoStop}%
\bibitem [{\citenamefont {Rakita}\ \emph {et~al.}(2017)\citenamefont {Rakita},
  \citenamefont {Bar-Elli}, \citenamefont {Meirzadeh}, \citenamefont {Kaslasi},
  \citenamefont {Peleg}, \citenamefont {Hodes}, \citenamefont {Lubomirsky},
  \citenamefont {Oron}, \citenamefont {Ehre},\ and\ \citenamefont
  {Cahen}}]{PNAS_non_centro}%
  \BibitemOpen
  \bibfield  {author} {\bibinfo {author} {\bibfnamefont {Y.}~\bibnamefont
  {Rakita}}, \bibinfo {author} {\bibfnamefont {O.}~\bibnamefont {Bar-Elli}},
  \bibinfo {author} {\bibfnamefont {E.}~\bibnamefont {Meirzadeh}}, \bibinfo
  {author} {\bibfnamefont {H.}~\bibnamefont {Kaslasi}}, \bibinfo {author}
  {\bibfnamefont {Y.}~\bibnamefont {Peleg}}, \bibinfo {author} {\bibfnamefont
  {G.}~\bibnamefont {Hodes}}, \bibinfo {author} {\bibfnamefont
  {I.}~\bibnamefont {Lubomirsky}}, \bibinfo {author} {\bibfnamefont
  {D.}~\bibnamefont {Oron}}, \bibinfo {author} {\bibfnamefont {D.}~\bibnamefont
  {Ehre}}, \ and\ \bibinfo {author} {\bibfnamefont {D.}~\bibnamefont {Cahen}},\
  }\href {\doibase 10.1073/pnas.1702429114} {\bibfield  {journal} {\bibinfo
  {journal} {Proceedings of the National Academy of Sciences}\ }\textbf
  {\bibinfo {volume} {114}},\ \bibinfo {pages} {E5504} (\bibinfo {year}
  {2017})}\BibitemShut {NoStop}%
\bibitem [{\citenamefont {G}\ \emph {et~al.}(2016)\citenamefont {G},
  \citenamefont {Mahale}, \citenamefont {Kore}, \citenamefont {Mukherjee},
  \citenamefont {Pavan}, \citenamefont {De}, \citenamefont {Ghara},
  \citenamefont {Sundaresan}, \citenamefont {Pandey}, \citenamefont
  {Guru~Row},\ and\ \citenamefont {Sarma}}]{DDsarma}%
  \BibitemOpen
  \bibfield  {author} {\bibinfo {author} {\bibfnamefont {S.}~\bibnamefont {G}},
  \bibinfo {author} {\bibfnamefont {P.}~\bibnamefont {Mahale}}, \bibinfo
  {author} {\bibfnamefont {B.~P.}\ \bibnamefont {Kore}}, \bibinfo {author}
  {\bibfnamefont {S.}~\bibnamefont {Mukherjee}}, \bibinfo {author}
  {\bibfnamefont {M.~S.}\ \bibnamefont {Pavan}}, \bibinfo {author}
  {\bibfnamefont {C.}~\bibnamefont {De}}, \bibinfo {author} {\bibfnamefont
  {S.}~\bibnamefont {Ghara}}, \bibinfo {author} {\bibfnamefont
  {A.}~\bibnamefont {Sundaresan}}, \bibinfo {author} {\bibfnamefont
  {A.}~\bibnamefont {Pandey}}, \bibinfo {author} {\bibfnamefont {T.~N.}\
  \bibnamefont {Guru~Row}}, \ and\ \bibinfo {author} {\bibfnamefont {D.~D.}\
  \bibnamefont {Sarma}},\ }\href {\doibase 10.1021/acs.jpclett.6b00803}
  {\bibfield  {journal} {\bibinfo  {journal} {The Journal of Physical Chemistry
  Letters}\ }\textbf {\bibinfo {volume} {7}},\ \bibinfo {pages} {2412}
  (\bibinfo {year} {2016})}\BibitemShut {NoStop}%
\bibitem [{\citenamefont {Lai}\ \emph {et~al.}(2018)\citenamefont {Lai},
  \citenamefont {Obliger}, \citenamefont {Lu}, \citenamefont {Kley},
  \citenamefont {Bischak}, \citenamefont {Kong}, \citenamefont {Lei},
  \citenamefont {Dou}, \citenamefont {Ginsberg}, \citenamefont {Limmer},\ and\
  \citenamefont {Yang}}]{soft}%
  \BibitemOpen
  \bibfield  {author} {\bibinfo {author} {\bibfnamefont {M.}~\bibnamefont
  {Lai}}, \bibinfo {author} {\bibfnamefont {A.}~\bibnamefont {Obliger}},
  \bibinfo {author} {\bibfnamefont {D.}~\bibnamefont {Lu}}, \bibinfo {author}
  {\bibfnamefont {C.~S.}\ \bibnamefont {Kley}}, \bibinfo {author}
  {\bibfnamefont {C.~G.}\ \bibnamefont {Bischak}}, \bibinfo {author}
  {\bibfnamefont {Q.}~\bibnamefont {Kong}}, \bibinfo {author} {\bibfnamefont
  {T.}~\bibnamefont {Lei}}, \bibinfo {author} {\bibfnamefont {L.}~\bibnamefont
  {Dou}}, \bibinfo {author} {\bibfnamefont {N.~S.}\ \bibnamefont {Ginsberg}},
  \bibinfo {author} {\bibfnamefont {D.~T.}\ \bibnamefont {Limmer}}, \ and\
  \bibinfo {author} {\bibfnamefont {P.}~\bibnamefont {Yang}},\ }\href {\doibase
  10.1073/pnas.1812718115} {\bibfield  {journal} {\bibinfo  {journal}
  {Proceedings of the National Academy of Sciences}\ }\textbf {\bibinfo
  {volume} {115}},\ \bibinfo {pages} {11929} (\bibinfo {year}
  {2018})}\BibitemShut {NoStop}%
\bibitem [{\citenamefont {Chen}\ \emph {et~al.}(2019)\citenamefont {Chen},
  \citenamefont {Luo}, \citenamefont {Fu}, \citenamefont {Wang}, \citenamefont
  {Czech}, \citenamefont {Shen}, \citenamefont {Guo}, \citenamefont {Wright},
  \citenamefont {Pan},\ and\ \citenamefont {Jin}}]{expt_strain}%
  \BibitemOpen
  \bibfield  {author} {\bibinfo {author} {\bibfnamefont {J.}~\bibnamefont
  {Chen}}, \bibinfo {author} {\bibfnamefont {Z.}~\bibnamefont {Luo}}, \bibinfo
  {author} {\bibfnamefont {Y.}~\bibnamefont {Fu}}, \bibinfo {author}
  {\bibfnamefont {X.}~\bibnamefont {Wang}}, \bibinfo {author} {\bibfnamefont
  {K.~J.}\ \bibnamefont {Czech}}, \bibinfo {author} {\bibfnamefont
  {S.}~\bibnamefont {Shen}}, \bibinfo {author} {\bibfnamefont {L.}~\bibnamefont
  {Guo}}, \bibinfo {author} {\bibfnamefont {J.~C.}\ \bibnamefont {Wright}},
  \bibinfo {author} {\bibfnamefont {A.}~\bibnamefont {Pan}}, \ and\ \bibinfo
  {author} {\bibfnamefont {S.}~\bibnamefont {Jin}},\ }\href {\doibase
  10.1021/acsenergylett.9b00543} {\bibfield  {journal} {\bibinfo  {journal}
  {ACS Energy Letters}\ }\textbf {\bibinfo {volume} {4}},\ \bibinfo {pages}
  {1045} (\bibinfo {year} {2019})}\BibitemShut {NoStop}%
\bibitem [{\citenamefont {Jishi}\ \emph {et~al.}(2014)\citenamefont {Jishi},
  \citenamefont {Ta},\ and\ \citenamefont {Sharif}}]{new_mBJ}%
  \BibitemOpen
  \bibfield  {author} {\bibinfo {author} {\bibfnamefont {R.~A.}\ \bibnamefont
  {Jishi}}, \bibinfo {author} {\bibfnamefont {O.~B.}\ \bibnamefont {Ta}}, \
  and\ \bibinfo {author} {\bibfnamefont {A.~A.}\ \bibnamefont {Sharif}},\
  }\href {\doibase 10.1021/jp5050145} {\bibfield  {journal} {\bibinfo
  {journal} {The Journal of Physical Chemistry C}\ }\textbf {\bibinfo {volume}
  {118}},\ \bibinfo {pages} {28344} (\bibinfo {year} {2014})}\BibitemShut
  {NoStop}%
\bibitem [{\citenamefont {Traor\'e}\ \emph {et~al.}(2019)\citenamefont
  {Traor\'e}, \citenamefont {Bouder}, \citenamefont {Lafargue-Dit-Hauret},
  \citenamefont {Rocquefelte}, \citenamefont {Katan}, \citenamefont {Tran},\
  and\ \citenamefont {Kepenekian}}]{new_mbj1}%
  \BibitemOpen
  \bibfield  {author} {\bibinfo {author} {\bibfnamefont {B.}~\bibnamefont
  {Traor\'e}}, \bibinfo {author} {\bibfnamefont {G.}~\bibnamefont {Bouder}},
  \bibinfo {author} {\bibfnamefont {W.}~\bibnamefont {Lafargue-Dit-Hauret}},
  \bibinfo {author} {\bibfnamefont {X.}~\bibnamefont {Rocquefelte}}, \bibinfo
  {author} {\bibfnamefont {C.}~\bibnamefont {Katan}}, \bibinfo {author}
  {\bibfnamefont {F.}~\bibnamefont {Tran}}, \ and\ \bibinfo {author}
  {\bibfnamefont {M.}~\bibnamefont {Kepenekian}},\ }\href {\doibase
  10.1103/PhysRevB.99.035139} {\bibfield  {journal} {\bibinfo  {journal} {Phys.
  Rev. B}\ }\textbf {\bibinfo {volume} {99}},\ \bibinfo {pages} {035139}
  (\bibinfo {year} {2019})}\BibitemShut {NoStop}%
\end{thebibliography}

\end{document}